\documentclass[pdflatex,sn-vancouver-num]{sn-jnl}

\usepackage{graphicx}%
\usepackage{multirow}%
\usepackage{amsmath,amssymb,amsfonts}%
\usepackage{amsthm}%
\usepackage{mathrsfs}%
\usepackage{xcolor}%
\usepackage{textcomp}%
\usepackage{manyfoot}%
\usepackage{booktabs}%
\usepackage{algorithm}%
\usepackage{algorithmicx}%
\usepackage{algpseudocode}%
\usepackage{listings}%
\usepackage{array}%
\usepackage{longtable}%
\usepackage{subcaption}%
\usepackage{mathtools}%
\usepackage{float}%
\usepackage[normalem]{ulem}%
\usepackage{cancel}%

\newcommand{\rev}[1]{#1}

\begin{document}

\title[A hybrid UQ toolbox]{A Hybrid Framework for Uncertainty Quantification in Partially Observed Dynamic Biological Systems}

\author[1]{\fnm{Alberto} \sur{Portela}}

\author*[1]{\fnm{Julio R.} \sur{Banga}}\email{j.r.banga@csic.es}

\affil*[1]{\orgdiv{Computational Biology Lab}, \orgname{MBG-CSIC (Spanish National Research Council)}, \orgaddress{\city{Pontevedra}, \country{Spain}}}
\abstract{
\textbf{Background.}
Mechanistic ordinary differential equation (ODE) models are widely used in systems biology, but uncertainty quantification (UQ) remains difficult when only a subset of state variables is experimentally observed. Existing Bayesian and likelihood-based approaches can be computationally demanding for nonlinear, weakly identifiable, or high-dimensional systems.

\textbf{Results.}
We present a framework, and its corresponding software \texttt{CUQDyn1\_Plus}, for UQ in partially observed ODE systems. Our method combines leave-one-out jackknife$+$-style empirical calibration for observed states with sensitivity-based Gaussian uncertainty propagation for hidden states. The software supports global parameter estimation, covariance propagation, bootstrap trajectory uncertainty and simulation-based calibration. It also facilitates comparison with Bayesian workflows, automated reporting, and reproducibility diagnostics. Validation on six benchmark systems shows accurate behavior in well-conditioned cases and model-dependent degradation under nonlinearity, weak identifiability, or global branch-switching non-identifiability.

\textbf{Conclusions.}
\texttt{CUQDyn1\_Plus} provides a practical and computationally efficient UQ workflow for systems biology models with observed and latent states. Its diagnostic outputs help identify when local Gaussian propagation is reliable and when uncertainty bands should be interpreted cautiously, making it a useful complement to fully Bayesian workflows.
}


\maketitle
\section{Introduction}
\label{sec:intro}

Mechanistic dynamic models formulated as systems of ordinary differential equations are central tools in systems biology, biotechnology, and biomedical engineering. These models provide quantitative descriptions of nonlinear biological processes and enable predictive analysis, hypothesis generation, model-based experimental design, and treatment optimization \citep{Braniff2018,Frhlich2018,mcdonald2023computational,banga2025,Goldring2025}. Despite their utility, uncertainty quantification (UQ) remains one of the major unresolved challenges in the practical use of mechanistic models \citep{mitra2019parameter,Sharp2022,villaverde2023assessment,qiao2025,BalsaCanto2025}.

Biological systems are typically only partially observed: measurements are noisy and sparse, model parameters are often poorly identifiable, and many biologically important state variables cannot be measured directly. Uncertainty in parameter estimates therefore propagates into model predictions for both observed and latent states, and reliable characterization of this uncertainty is required before model predictions can be used for decision-making \citep{Mikovi2010,Kaltenbach2009,kirk2015systems, cedersund2016prediction,banga2025}.

A wide range of approaches has been proposed for UQ in dynamic systems. Bayesian Markov Chain Monte Carlo (MCMC) methods naturally provide full posterior uncertainty distributions, but can become computationally prohibitive for high-dimensional nonlinear systems and may suffer from convergence failures in the presence of non-identifiability or multimodal likelihood surfaces \citep{raue2013joining,Liepe2014,hines2014determination}. Profile likelihood methods provide strong inferential foundations but can scale poorly when trajectory-level uncertainty must be computed repeatedly across many prediction points \citep{kreutz2013profile, villaverde2023assessment}. Fisher Information Matrix (FIM)-based local approximations offer computational efficiency but rely on quadratic likelihood approximations that may be unreliable for nonlinear or poorly identifiable models \citep{villaverde2023assessment}. Ensemble methods can improve scalability at the cost of weaker theoretical justification \citep{massonis2023improving}.

In a recent study, we introduced conformal prediction algorithms for uncertainty quantification in fully observed dynamical systems \citep{portela2025conformal}. Those methods leveraged leave-one-out (LOO) conformal prediction to provide finite-sample predictive guarantees for observed state trajectories without requiring strong parametric assumptions. Across several benchmark systems of increasing complexity the conformal algorithms proved substantially faster than Bayesian MCMC approaches while achieving comparable or better calibration. For large-scale systems where Bayesian tools encountered convergence failures, the conformal methods produced valid prediction regions within minutes on a standard workstation.

However, the original conformal framework was restricted to observed variables. In many practical systems biology applications, only a subset of the model states are experimentally measurable. Hidden states may nonetheless be biologically important quantities: the predator population in an ecosystem, latent infection compartments in an epidemiological model, unmeasured nuclear signaling species, or unobservable intermediate metabolites. Extending conformal prediction directly to hidden states is fundamentally limited because no direct calibration residuals are available for variables that are never observed.

To address this limitation, in this study we develop a hybrid conformal--Gaussian framework for partially observed ODE systems and introduce the MATLAB toolbox \texttt{CUQDyn1\_Plus}. The key organizing principle is to treat observed and unobserved states asymmetrically, because they admit fundamentally different statistical treatments. \rev{For observed states, the toolbox uses a leave-one-out jackknife$+$-style empirical calibration based on held-out residuals \cite{barber2021predictive,portela2025conformal}. We describe this as conformal-style calibration because it follows the jackknife$+$ construction, while noting that exact finite-sample guarantees depend on exchangeability and order-statistic details.} For unobserved states, no direct calibration is possible, and uncertainty must be propagated analytically through the ODE system using parameter covariance estimates and trajectory sensitivity matrices.


The toolbox further supports automated problem definition, synthetic data generation, LOO diagnostics, covariance condition number assessment, and near-bound parameter diagnostics. We also provide Bayesian comparison workflows through a companion \texttt{PyMC} \cite{pymc2023} implementation using approximate Bayesian computation sequential Monte Carlo (ABC-SMC). \texttt{PyMC} provides a modern probabilistic programming framework in Python \cite{pymc2023,helleckes2022bayesian,castro2025approximate}. Our workflow further extends its usability by offering seamless integration with \texttt{CUQDyn1\_Plus}, facilitating comparative analyses and model evaluation across both environments.

The main contributions of this work are:
\begin{enumerate}
    \item A \textbf{hybrid conformal--Gaussian framework} for UQ in partially observed
    ODE systems, combining jackknife$+$-style empirical conformal bands for
    observed states with delta-method Gaussian propagation for latent states.
    \item A \textbf{hybrid covariance strategy} (\texttt{HybridCov}) that pairs FIM
    marginal variance scales with empirical leave-one-out parameter correlations,
    \rev{providing an alternative latent-state covariance construction whose benefit
    is assessed empirically against FIM-only and naive LOO-only covariance approaches.}
    \item The \textbf{\texttt{CUQDyn1\_Plus} MATLAB toolbox}, implementing all three
    UQ workflows with global parameter estimation, parallelized LOO ensemble
    construction, complex-step sensitivity analysis, automated diagnostics, and
    standardized output reporting.
    \item \textbf{Simulation-based calibration} experiments across benchmark systems
    that provide direct empirical evidence on coverage properties and reveal
    system-dependent limitations of the local Gaussian approximation.
    \item A \textbf{systematic comparison with Bayesian inference} via a companion
    \texttt{PyMC} ABC-SMC implementation, demonstrating conditions under which each approach
    is preferable.
\end{enumerate}

The remainder of this manuscript is organized as follows. Section~\ref{sec:methods} introduces the mathematical framework and uncertainty quantification methodology. Section~\ref{sec:examples} presents representative benchmark examples. Section~\ref{sec:pymc_comparison} compares our results with Bayesian inference. Section~\ref{sec:discussion} discusses the strengths, limitations, and future directions of the proposed framework. Additional file 1 provides supplementary run settings, generated result tables, SBC summaries, identifiability diagnostics, software implementation details, and CUQDyn1\_Plus--PyMC comparison results.

\section{Methods}
\label{sec:methods}

\subsection{Modeling framework and notation}
\label{sec:model}

We consider deterministic nonlinear ODE systems of the form
\begin{align}
\dot{x}(t) &= f(x(t), \theta, t), \label{eq:ode}\\
x(t_0) &= x_0, \label{eq:ic}\\
y(t) &= g(x(t), \theta, t), \label{eq:obs}
\end{align}
where $x(t) \in \mathbb{R}^{n_x}$ denotes the vector of state variables, $y(t) \in \mathbb{R}^{n_{\rm out}}$ denotes the vector of model outputs, and $\theta \in \mathbb{R}^{n_\theta}$ denotes the vector of unknown parameters. The vector field $f : \mathbb{R}^{n_x} \times \mathbb{R}^{n_\theta} \times \mathbb{R} \to \mathbb{R}^{n_x}$ and the observation mapping $g : \mathbb{R}^{n_x} \times \mathbb{R}^{n_\theta} \times \mathbb{R} \to \mathbb{R}^{n_{\rm out}}$ are generally nonlinear. \rev{For the benchmark examples, $g$ is usually the identity or a state-selection map, but the toolbox stores explicit observed-state indices so that the number of model states, model outputs, and measured outputs need not be confused.}

A key feature of the setting considered here is that only a subset of the $n_x$ states is experimentally observed after the initial condition. Let \(\mathcal{O}\subset\{1,\ldots,n_x\}\) denote the directly observed state-index set, with cardinality \(n_{\rm obs}=|\mathcal{O}|\), and let \(\mathcal{H}\) denote its complement. We write 
\(x_{\mathcal{O}}(t)\) for directly observed states and \(x_{\mathcal{H}}(t)\) for latent states. States in \(\mathcal{H}\) are not directly measured after the initial time. The initial condition provides values for all states at \(t_0\).

Observed measurements are modeled through an observation map
\(g_k(x(t),\theta,t)\) with additive measurement noise,
\begin{equation}
\tilde{y}_{k,i} = g_k(x(t_i,\theta),\theta,t_i) + \epsilon_{k,i},
\quad i = 1, \ldots, m;\; k = 1, \ldots, n_{\rm obs},
\label{eq:noise}
\end{equation}
where \(\epsilon_{k,i}\) denotes additive zero-mean measurement noise for the
\(k\)-th observed output. In the benchmark systems considered here, the
observation map corresponds to direct noisy measurements of selected state
variables, so \(g_k(x(t),\theta,t)=x_{o_k}(t,\theta)\) for
\(o_k\in\mathcal{O}\). For parameter estimation, we assume a Gaussian noise
model \(\epsilon_{k,i} \sim \mathcal{N}(0,\sigma_k^2)\), where \(\sigma_k\) is
the standard deviation for that observed output. The toolbox supports known
measurement variances provided by the user, variances estimated from residuals,
and state- or output-specific weighting for situations where observed quantities
have substantially different physical scales or measurement accuracies.

\subsection{Global parameter estimation}
\label{sec:estimation}

Parameter estimation is formulated as a weighted nonlinear least-squares optimization problem:
\begin{equation}
\hat{\theta} = \arg\min_{\theta \in \Theta} \; \sum_{k=1}^{n_{\rm obs}} \sum_{i=1}^{m} \left( \frac{\tilde{y}_{k,i} - x_{o_k}(t_i, \theta)}{\sigma_k} \right)^2,
\label{eq:lsq}
\end{equation}
where $\Theta = [\theta_{\min}, \theta_{\max}]$ denotes the parameter feasible set defined by lower and upper bounds. The optimization is carried out using a global optimizer, enhanced scatter search (eSS), part of the MEIGO toolbox \citep{egea2014meigo}. The eSS strategy combines stochastic global search with deterministic local improvement and has demonstrated robust performance on challenging nonlinear parameter estimation problems in systems biology \citep{villaverde2019benchmarking}.

Three residual weighting models are supported: unweighted residuals, known measurement standard deviations, and user-specified state weights. The residual model affects both the optimization objective and the subsequent FIM covariance estimation, and should be chosen consistently to reflect the assumed measurement error structure.

The full-data parameter estimate $\hat{\theta}$ and the corresponding fitted trajectory $\hat{x}(t) = x(t, \hat{\theta})$ form the starting point for all downstream uncertainty quantification steps (Algorithm~\ref{alg:full_fit}).

\begin{algorithm}[H]
\caption{Full-data parameter estimation}
\label{alg:full_fit}
\begin{algorithmic}[1]
\Require times $\{t_i\}_{i=0}^{m}$, observations $\tilde{y}_{k,i}$, initial state $x_0$, bounds $\Theta$, initial guess $\theta_0$, residual model
\Ensure full-data estimate $\hat{\theta}$ and fitted trajectory $\hat{x}(t_i)$
\State Define weighted residuals
\[
R_{k,i}(\theta) =
\begin{cases}
\tilde{y}_{k,i} - x_{o_k}(t_i,\theta), & \texttt{none},\\
(\tilde{y}_{k,i} - x_{o_k}(t_i,\theta))/\sigma_k, & \texttt{known\_sigma},\\
w_k(\tilde{y}_{k,i} - x_{o_k}(t_i,\theta)), & \texttt{state\_weights}.
\end{cases}
\]
\State Solve
\[
\hat{\theta}
=
\arg\min_{\theta \in \Theta}
\sum_{i=1}^{m}\sum_{k=1}^{n_{\rm obs}} R_{k,i}(\theta)^2
\]
using MEIGO/eSS.
\State Integrate $\dot{x}=f(x,\hat{\theta},t)$ from $x_0$ to obtain $\hat{x}(t_i)$.
\State \Return $\hat{\theta}$, $\hat{x}(t_i)$.
\end{algorithmic}
\end{algorithm}

\subsection{Leave-one-out ensemble}
\label{sec:loo}

The conformal and hybrid uncertainty quantification methods are built on a leave-one-out (LOO) ensemble of model refits. Let $\{t_1, \ldots, t_m\}$ denote the $m$ observed time points for the dataset (excluding the initial condition at $t_0$). For each index $i \in \{1, \ldots, m\}$, the model is refitted to all observations except the $i$-th time point:
\begin{equation}
\hat{\theta}_{(-i)} = \arg\min_{\theta \in \Theta} \; \sum_{k=1}^{n_{\rm obs}} \sum_{j \neq i} \left( \frac{\tilde{y}_{k,j} - x_{o_k}(t_j, \theta)}{\sigma_k} \right)^2.
\label{eq:loo_fit}
\end{equation}

This produces $m$ LOO parameter vectors $\{\hat{\theta}_{(-i)}\}_{i=1}^m$, $m$ corresponding ODE trajectory solutions $\{\hat{x}(t, \hat{\theta}_{(-i)})\}_{i=1}^m$, and $m$ held-out absolute prediction residuals:
\begin{equation}
r_{k,i} = \left| \tilde{y}_{k,i} - x_{o_k}(t_i, \hat{\theta}_{(-i)}) \right|, \quad k = 1, \ldots, n_{\rm obs}.
\label{eq:loo_resid}
\end{equation}
The LOO ensemble $\{\hat{\theta}_{(-i)}\}$ and the held-out residuals $\{r_{k,i}\}$ are the two central objects used by the conformal calibration and hybrid covariance construction described in the following sections (Algorithm~\ref{alg:loo}).

LOO refits are parallelized over time points using MATLAB \texttt{parfor} loops, substantially reducing wall-clock time on multi-core workstations. Each LOO refit can be performed with either global eSS optimization, which is the default and is robust but expensive, or a guarded local--after--global strategy that initializes from $\hat{\theta}$ and falls back to global search if the local fit is unsatisfactory. The local-after-global strategy is substantially faster and is used by selected fast-SBC workflows, with diagnostic checks stored to detect local failures. The bootstrap routine performs repeated MEIGO refits by default unless a custom fit handle is supplied.

\begin{algorithm}[H]
\caption{Leave-one-out refit ensemble}
\label{alg:loo}
\begin{algorithmic}[1]
\Require full dataset, full-data estimate $\hat{\theta}$, bounds $\Theta$, initial state $x_0$
\Ensure LOO parameters $\{\hat{\theta}_{(-i)}\}_{i=1}^{m}$, LOO trajectories, held-out residuals
\For{$i=1,\ldots,m$}
    \State Remove post-initial observation time $t_i$ from the fitting data.
    \State Refit the model:
    \[
    \hat{\theta}_{(-i)}
    =
    \arg\min_{\theta \in \Theta}
    \sum_{j\neq i}\sum_{k=1}^{n_{\rm obs}} R_{k,j}(\theta)^2 .
    \]
    \State Integrate $\dot{x}=f(x,\hat{\theta}_{(-i)},t)$ at all times.
    \State Store held-out residuals
    \[
    r_{k,i}
    =
    \left|\tilde{y}_{k,i}
    -
    x_{o_k}(t_i,\hat{\theta}_{(-i)})\right|.
    \]
\EndFor
\State \Return $\{\hat{\theta}_{(-i)}\}$, LOO trajectories, $\{r_{k,i}\}$.
\end{algorithmic}
\end{algorithm}

\subsection{Observed-state conformal prediction}
\label{sec:conformal}

For states that are directly observed, the toolbox computes prediction intervals using the LOO conformal framework introduced in \citet{portela2025conformal}, adapted to the partially observed setting. The approach is a variant of the jackknife$+$ method of \citet{barber2021predictive}.

For observed-state column $k$ with state index $o_k\in\mathcal{O}$ and time point $i$, let $\mathcal{E}_{k,i} = (x_{o_k}(t_i, \hat{\theta}_{(-1)}), \ldots, x_{o_k}(t_i, \hat{\theta}_{(-m)}))$ denote the vector of LOO ensemble predictions at that state and time, and let $\mathbf{r}_k = (r_{k,1}, \ldots, r_{k,m})$ denote the corresponding vector of held-out absolute residuals across all time points for observed-state column $k$. The lower and upper conformal prediction bands are computed as
\begin{align}
L_{k,i} &= Q_{\alpha}(\mathcal{E}_{k,i} - \mathbf{r}_k), \label{eq:conf_lower}\\
U_{k,i} &= Q_{1-\alpha}(\mathcal{E}_{k,i} + \mathbf{r}_k), \label{eq:conf_upper}
\end{align}
where $Q_\alpha(\cdot)$ denotes the empirical $\alpha$-quantile of the argument vector, and $\alpha \in (0, 0.5)$ controls the nominal coverage level $1 - 2\alpha$. With $\alpha = 0.025$ the nominal two-sided coverage is 95\%.

The construction follows the jackknife$+$ principle: the prediction interval at each time point is formed by perturbing the LOO ensemble predictions by the LOO residuals and taking empirical quantiles. \rev{The current MATLAB implementation uses the empirical quantile operator, so the bands should be interpreted as jackknife$+$-style conformal empirical intervals. Exact finite-sample jackknife$+$ guarantees require the exchangeability assumptions of \citet{barber2021predictive} and an order-statistic calibration rule rather than interpolated quantiles. Under those assumptions, the target marginal guarantee for a new exchangeable observation is} (Algorithm~\ref{alg:conformal}):
\begin{equation}
\mathbb{P}\!\left(\tilde{y}_{k,n+1} \in [L_{k,n+1}, U_{k,n+1}]\right) \geq 1 - 2\alpha,
\label{eq:coverage_guarantee}
\end{equation}
where the probability is over the joint distribution of training and test data \citep{barber2021predictive}. \rev{The robustness property is therefore best described as conformal-style, distribution-light calibration rather than an unconditional guarantee for arbitrary ODE time series. It does not require Gaussian residuals or a correctly specified noise model, but it does require an exchangeability approximation; empirical coverage should be checked with diagnostics or SBC experiments.}

\begin{algorithm}[H]
\caption{Observed-state jackknife+ conformal bands}
\label{alg:conformal}
\begin{algorithmic}[1]
\Require LOO trajectories, held-out residuals $r_{k,i}$, tail probability $\alpha$
\Ensure conformal lower and upper bands $L_{k,i}$ and $U_{k,i}$ for observed states
\For{each observed-state column $k=1,\ldots,n_{\rm obs}$, with state index $o_k$}
    \State Form residual vector $\mathbf{r}_k=(r_{k,1},\ldots,r_{k,m})$.
    \For{each post-initial time $t_i$}
        \State Form LOO prediction vector
        \[
        \mathcal{E}_{k,i}
        =
        \left(
        x_{o_k}(t_i,\hat{\theta}_{(-1)}),
        \ldots,
        x_{o_k}(t_i,\hat{\theta}_{(-m)})
        \right).
        \]
        \State Compute
        \[
        L_{k,i}=Q_{\alpha}(\mathcal{E}_{k,i}-\mathbf{r}_k),
        \qquad
        U_{k,i}=Q_{1-\alpha}(\mathcal{E}_{k,i}+\mathbf{r}_k).
        \]
    \EndFor
\EndFor
\State Set the initial-time band equal to the known initial condition.
\State \Return $L_{k,i}$, $U_{k,i}$.
\end{algorithmic}
\end{algorithm}

\subsection{FIM-based latent-state uncertainty propagation}
\label{sec:fim}

For state variables that are never directly observed, no calibration residuals are available, and the conformal construction of Section~\ref{sec:conformal} cannot be directly applied. Uncertainty for these latent states must instead be propagated analytically from the estimated parameter uncertainty to the ODE trajectory.

The parameter covariance is estimated through the Gauss-Newton approximation to the Fisher Information Matrix \citep{bard1974nonlinear,villaverde2023assessment}. \rev{In the current toolbox this computation is performed by a shared helper which uses log-parameter coordinates by default. Let $\phi=\log\theta$ and let $J_\phi$ be the Jacobian of the weighted residual vector with respect to $\phi$, evaluated at $\hat{\phi}=\log\hat{\theta}$. The log-space covariance is regularized in a scale-relative way,}
\begin{equation}
\rev{
\text{Cov}_{\phi}
=
\hat{\sigma}^2
\left(J_\phi^T J_\phi + \lambda_{\mathrm{rel}}\max(\mu_{\max},1) I\right)^{-1},
\qquad
\mu_{\max}=\lambda_{\max}(J_\phi^T J_\phi),
}
\label{eq:fim_cov}
\end{equation}
\rev{where $\lambda_{\mathrm{rel}}$ can be controlled by the user; the $\max(\mu_{\max},1)$ factor matches the scale-relative ridge used in the implementation. The covariance returned to the rest of the toolbox is transformed back to natural parameter coordinates,}
\begin{equation}
\rev{
\text{Cov}_{\theta}
=
\operatorname{diag}(\hat{\theta})\,\text{Cov}_{\phi}\,\operatorname{diag}(\hat{\theta}) .
}
\label{eq:fim_cov_natural}
\end{equation}
\rev{The helper also computes an SVD of $J_\phi$ to estimate numerical rank and condition number, identifies hard rank-deficient directions under the configured tolerance, and records the fraction of each state sensitivity lying in those weak directions.} All entries of the residual Jacobian are computed using complex-step differentiation \citep{martins2003complex}, which provides near-machine-precision derivative estimates and avoids the step-size sensitivity and cancellation errors associated with real-valued finite differences. \rev{When known measurement standard deviations are supplied, the residual variance scale is set to one; otherwise it is estimated from the fitted residuals using the post-initial observation rows.}

Trajectory prediction uncertainty is propagated from parameter space to state space using the delta method (first-order linearization):
\begin{equation}
\text{Var}(x_k(t)) \approx S_k(t) \; \text{Cov}_{\theta} \; S_k(t)^T,
\label{eq:delta_var}
\end{equation}
where
\begin{equation}
S_k(t) = \frac{\partial x_k(t, \theta)}{\partial \theta}\bigg|_{\theta = \hat{\theta}} \in \mathbb{R}^{1 \times n_\theta}
\label{eq:sensitivity}
\end{equation}
is the trajectory sensitivity vector for state $k$ at time $t$, computed by complex-step ODE solves. The prediction interval for an unobserved state is then
\begin{equation}
\hat{x}_k(t) \pm z_{1-\alpha} \sqrt{\text{Var}(x_k(t))},
\label{eq:fim_band}
\end{equation}
where $z_{1-\alpha}$ is the $(1-\alpha)$ quantile of the standard normal distribution. Note that \(z_{1-\alpha}\) is the standard-normal quantile, with \(\alpha\) again
interpreted as the one-sided tail probability. The complete procedure is given in Algorithm~\ref{alg:fim_delta}.

The FIM delta-method approach is fast and interpretable but relies on local linearity of the ODE trajectories with respect to parameters near $\hat{\theta}$, approximate Gaussianity of parameter uncertainty, and reasonably accurate covariance estimation from the Gauss-Newton approximation. These assumptions are most reliable when the model is locally identifiable, the number of parameters is not too large relative to the data, and the ODE trajectories vary smoothly with parameters. \rev{For weakly identifiable systems, the numerical covariance is necessarily regularization-dependent. The current implementation therefore treats the SVD rank and weak-direction sensitivity fractions as first-class diagnostics rather than hiding conditioning problems inside a single covariance matrix.}

\begin{algorithm}[H]
\caption{FIM delta-method bands for latent states}
\label{alg:fim_delta}
\begin{algorithmic}[1]
\Require full-data estimate $\hat{\theta}$, fitted trajectory $\hat{x}(t_i)$, weighted residual function $R(\theta)$, tail probability $\alpha$
\Ensure Gaussian delta-method bands for unobserved states
\State \rev{Compute the natural-parameter residual Jacobian by complex-step differentiation; the shared helper rescales it to the selected FIM parameterization, log space by default:}
\[
\rev{J_\theta=\left.\frac{\partial R(\theta)}{\partial\theta}\right|_{\theta=\hat{\theta}}, \qquad J_\phi=J_\theta\operatorname{diag}(\hat{\theta}) .}
\]
\State Estimate the residual variance scale $\hat{\sigma}^2$; set $\hat{\sigma}^2=1$ when known measurement standard deviations are supplied.
\State \rev{Compute the singular values of $J_\phi$ and store rank/conditioning diagnostics. Form}
\[
\rev{\text{Cov}_{\phi}}
=
\hat{\sigma}^2
\rev{\left(J_\phi^T J_\phi+\lambda_{\mathrm{rel}}\max(\mu_{\max},1) I\right)^{-1}.}
\]
\State \rev{Transform to natural parameter coordinates: $\text{Cov}_{\theta}=\operatorname{diag}(\hat{\theta})\text{Cov}_{\phi}\operatorname{diag}(\hat{\theta})$.}
\State Compute natural-parameter trajectory sensitivities by complex-step ODE solves; internally, the FIM helper rescales these sensitivities consistently when log-parameter diagnostics are used:
\[
S_k(t_i)
=
\left.
\frac{\partial x_k(t_i,\theta)}{\partial\theta}
\right|_{\theta=\hat{\theta}} .
\]
\For{each unobserved state $k$ and time $t_i$}
    \State Compute
    \[
    v_{k,i}
    =
    S_k(t_i)\text{Cov}_{\theta}S_k(t_i)^T .
    \]
    \State \rev{Record the state-level reliability diagnostic: whether the maximum weak-direction sensitivity fraction exceeds the configured threshold.}
    \State Set
    \[
    L_{k,i}=\hat{x}_k(t_i)-z_{1-\alpha}\sqrt{v_{k,i}},
    \qquad
    U_{k,i}=\hat{x}_k(t_i)+z_{1-\alpha}\sqrt{v_{k,i}}.
    \]
\EndFor
\State \Return $L_{k,i}$, $U_{k,i}$ for unobserved states.
\end{algorithmic}
\end{algorithm}

\subsection{Hybrid covariance strategy}
\label{sec:hybridcov}

Although the FIM approach often provides a reasonable estimate of the marginal parameter variances, it may produce unreliable parameter correlation structures in weakly identifiable or high-dimensional systems. The correlation structure encoded in $\text{Cov}_{\text{FIM}}$ is derived from a quadratic approximation to the likelihood surface, which can be inaccurate when the surface is curved, ridged, or multi-modal.

To improve the correlation estimate while preserving the FIM's well-calibrated marginal scales, we introduce a hybrid covariance estimator:
\begin{equation}
\text{Cov}_{\text{hyb}} = D_{\text{FIM}} \; R_{\text{LOO}} \; D_{\text{FIM}},
\label{eq:hybrid_cov}
\end{equation}
where $D_{\text{FIM}} = \text{diag}(\sqrt{\text{Cov}_{\theta,11}}, \ldots, \sqrt{\text{Cov}_{\theta,n_\theta n_\theta}})$ is the diagonal matrix of FIM marginal standard deviations in natural parameter coordinates, and $R_{\text{LOO}}$ is the empirical parameter correlation matrix estimated from the LOO parameter ensemble:
\begin{equation}
R_{\text{LOO}} = \text{corr}(\hat{\Theta}_{\text{LOO}}), \quad \hat{\Theta}_{\text{LOO}} = [\hat{\theta}_{(-1)}, \ldots, \hat{\theta}_{(-m)}]^T \in \mathbb{R}^{m \times n_\theta}.
\label{eq:rloo}
\end{equation}
The implementation symmetrizes $\text{Cov}_{\text{hyb}}$ and numerically projects it to a positive semi-definite matrix if small negative eigenvalues arise. Intuitively, each LOO refit removes one time point from the data and refits the model, thereby inducing a small but meaningful shift in the parameter estimates. The correlation structure of these shifts reflects the effective parameter interdependencies in the ODE model as constrained by the data, providing an empirical alternative to the quadratic FIM correlation, subject to the stability and informativeness of the LOO refits.

It is important to note that using the LOO covariance $\hat{\Sigma}_{\text{LOO}} = \text{Cov}(\hat{\Theta}_{\text{LOO}})$ directly as the parameter covariance matrix would severely underestimate the true estimation variance: since removing one of $m$ time points barely shifts the parameter optimum, the spread of $\hat{\Theta}_{\text{LOO}}$ is smaller than the true sampling variability of $\hat{\theta}$ by a factor of order $O(m)$. Simulation-based calibration experiments confirm this: the naive LOO-only covariance strategy is strongly under-dispersed. \rev{The hybrid strategy in Eq.~(\ref{eq:hybrid_cov}) addresses this by anchoring the marginal scales to the FIM estimate while retaining the empirical correlation structure from the LOO ensemble. Its performance should be reported as model-dependent: it can improve latent-state calibration when the empirical LOO correlation is informative, but shared-fit comparisons show that it is not uniformly superior to the FIM-only variant.}

After assembling $\text{Cov}_{\text{hyb}}$, the same delta-method propagation described in Section~\ref{sec:fim} is applied to obtain prediction bands for unobserved states (Algorithm~\ref{alg:hybridcov}). The observed-state conformal bands remain identical in both the FIM and HybridCov workflows.

\begin{algorithm}[H]
\caption{HybridCov latent-state bands}
\label{alg:hybridcov}
\begin{algorithmic}[1]
\Require FIM covariance $\text{Cov}_{\text{FIM}}$, LOO parameter matrix $\hat{\Theta}_{\text{LOO}}$, fitted trajectory $\hat{x}(t_i)$, tail probability $\alpha$
\Ensure HybridCov delta-method bands for unobserved states
\State Extract FIM marginal scales \rev{from the natural-space FIM covariance $\text{Cov}_{\theta}$}:
\[
D_{\text{FIM}}
=
\text{diag}
\left(
\sqrt{\text{Cov}_{\theta,11}},
\ldots,
\sqrt{\text{Cov}_{\theta,n_\theta n_\theta}}
\right).
\]
\State Estimate the LOO covariance and correlation:
\[
\hat{\Sigma}_{\text{LOO}}=\text{Cov}(\hat{\Theta}_{\text{LOO}}),
\qquad
R_{\text{LOO}}=\text{corr}(\hat{\Theta}_{\text{LOO}}).
\]
\State Assemble
\[
\text{Cov}_{\text{hyb}}
=
D_{\text{FIM}}R_{\text{LOO}}D_{\text{FIM}}.
\]
\State Symmetrize $\text{Cov}_{\text{hyb}}$ and project small negative eigenvalues to zero if necessary.
\State Propagate $\text{Cov}_{\text{hyb}}$ through the same trajectory sensitivities as in Algorithm~\ref{alg:fim_delta}.
\State Use observed-state conformal bands unchanged.
\State \Return full-state bands combining conformal observed-state bands and HybridCov latent-state bands.
\end{algorithmic}
\end{algorithm}

\subsection{Parametric bootstrap trajectory uncertainty}
\label{sec:bootstrap}

As an optional post-fit robustness check, the toolbox implements parametric bootstrap trajectory uncertainty analysis. The bootstrap workflow starts from an existing \texttt{CUQDyn1\_Plus} or \texttt{CUQDyn1\_Plus\_HybridCov} result and proceeds in five steps (Algorithm~\ref{alg:bootstrap}). First, the residual noise standard deviation is estimated from the post-initial observed-state residuals of the full-data fit. Second, $B$ synthetic observed datasets are generated by adding independent Gaussian draws, using the corresponding estimated standard deviation for each observed state, to the fitted observed-state trajectory while keeping the initial observed row fixed. Third, for each synthetic dataset $b = 1, \ldots, B$, the model is refitted to obtain a bootstrap parameter estimate $\hat{\theta}^{(b)}$. Fourth, the ODE is solved for each $\hat{\theta}^{(b)}$ to obtain a bootstrap trajectory ensemble. Fifth, empirical lower and upper quantile bands are computed across the $B$ bootstrap trajectories at each time point.

The resulting bootstrap bands are latent ODE trajectory bands, not noisy-observation prediction intervals. They characterize uncertainty in the deterministic ODE trajectory itself induced by parameter estimation uncertainty and the assumed bootstrap noise model. As such, they should not be compared directly with the spread of noisy observations when assessing coverage. The distinction between observed-data prediction intervals and latent trajectory bands is critical for correct interpretation, and the toolbox documentation and generated reports record this distinction explicitly.

Compared with the delta-method approaches, the bootstrap workflow requires considerably more computation because it performs $B$ complete model refits. However, it is less dependent on the local linearity and Gaussianity assumptions underlying the delta method, and can capture some nonlinear and asymmetric trajectory uncertainty. For publication-quality bootstrap summaries, a minimum of $B = 100$ replicates is recommended, with $B = 50$ used for tutorial purposes.

\begin{algorithm}[H]
\caption{Parametric bootstrap trajectory bands}
\label{alg:bootstrap}
\begin{algorithmic}[1]
\Require fitted result $(\hat{\theta},\hat{x}(t_i))$, observed data, bounds $\Theta$, number of replicates $B$, tail probability $\alpha$
\Ensure bootstrap trajectory bands for all states
\State Estimate one noise scale per observed state:
\[
\hat{\sigma}_k
=
\sqrt{
\frac{1}{m}
\sum_{i=1}^{m}
\left(\tilde{y}_{k,i}-\hat{x}_{o_k}(t_i)\right)^2
}.
\]
\For{$b=1,\ldots,B$}
    \State Generate bootstrap observations for post-initial times:
    \[
    \tilde{y}^{(b)}_{k,i}
    =
    \hat{x}_{o_k}(t_i)+\eta^{(b)}_{k,i},
    \qquad
    \eta^{(b)}_{k,i}\sim\mathcal{N}(0,\hat{\sigma}_k^2).
    \]
    \State Keep the initial observed row fixed.
    \State Refit the model to obtain $\hat{\theta}^{(b)}$.
    \State Integrate $\dot{x}=f(x,\hat{\theta}^{(b)},t)$ and store $x^{(b)}(t_i)$.
\EndFor
\For{each state $k$ and time $t_i$}
    \State Compute empirical quantiles:
    \[
    L_{k,i}=Q_\alpha\left(\{x^{(b)}_k(t_i)\}_{b=1}^{B}\right),
    \qquad
    U_{k,i}=Q_{1-\alpha}\left(\{x^{(b)}_k(t_i)\}_{b=1}^{B}\right).
    \]
\EndFor
\State \Return bootstrap trajectory bands $L_{k,i}$ and $U_{k,i}$.
\end{algorithmic}
\end{algorithm}

\subsection{Computational complexity}
\label{sec:complexity}

The dominant computational cost of the conformal workflow arises from the $m$ LOO refits. For a model with $m$ measurement time points, $n_\theta$ parameters, and per-refit optimization cost proportional to $k \cdot m \cdot n_\theta$ (where $k$ denotes the number of optimizer iterations), the overall LOO ensemble cost scales approximately as
\begin{equation}
\mathcal{O}(m^2 \cdot n_\theta \cdot k).
\label{eq:complexity}
\end{equation}
This quadratic scaling in $m$ is the main limitation of the LOO conformal approach for large datasets. However, for the sparse time series typical of systems biology experiments (often $n < 100$ observations), this scaling is manageable. Parallelization over LOO refits reduces wall-clock time approximately linearly with the number of available cores. FIM and sensitivity computations require repeated complex-step ODE solves: roughly one set for the residual Jacobian and one set for trajectory sensitivities, scaling approximately linearly with $n_\theta$, the number of time points, and the ODE solve cost.

In our original benchmark study \citep{portela2025conformal}, the conformal algorithms were up to two orders of magnitude faster than Bayesian MCMC (Stan) for equivalent problems. The Bayesian cost scales as $\mathcal{O}(S \cdot k \cdot m \cdot n_\theta)$ where $S$ may be $10^4$--$10^5$ posterior samples and $k$ is the number of leapfrog steps per HMC iteration, making MCMC considerably more expensive even for moderate-sized systems. The bootstrap workflow has cost $\mathcal{O}(B \cdot m \cdot n_\theta \cdot k)$ and is therefore intermediate between the LOO conformal workflow and full Bayesian sampling.

\subsection{Availability and Implementation}
\label{sec:availability}

\texttt{CUQDyn1\_Plus} is implemented in MATLAB and released under the \rev{GPLv3} license. The software, including instructions for running all experiments, is available at Zenodo: \href{https://doi.org/10.5281/zenodo.21470609}{https://doi.org/10.5281/zenodo.21470609}. This distribution includes source code, maintained benchmark examples, simulation-based calibration workflows, Bayesian comparison workflows and detailed documentation. 

\section{Benchmark Examples}
\label{sec:examples}

We validate the toolbox on six dynamical systems of increasing complexity, summarized in Table~\ref{tab:benchmarks}. Two linear cascade systems serve as analytic verification cases with known closed-form solutions; the remaining four are nonlinear biological benchmarks. Each system is treated as a partially observed problem: the initial condition is fully specified, but after $t_0$ only a designated subset of states is observed. Synthetic benchmark datasets are generated with additive Gaussian noise at 10\% of the mean trajectory amplitude, except for AP, where the comparison uses the experimental $y_1$--$y_4$ measurement dataset, and for NF-$\kappa$B, whose maintained synthetic dataset uses 5\% noise over 36 sampling times with the specified ten observed-state subset.

In all examples, \(n_x\) denotes the dimension of the ODE state vector and
\(n_{\rm obs}\) denotes the number of state variables directly measured after the
initial condition. Several analyses also report uncertainty bands for unobserved
state variables. Thus, the number of states displayed in figures or exported in
summary tables may be larger than \(n_{\rm obs}\), because latent-state uncertainty
is propagated from the fitted parameter uncertainty even when those states are not
directly measured.

The diagnostic cascade examples observe only the terminal species. The
Lotka--Volterra and SIR examples use one observed state to assess recovery of the
remaining latent state variables. The $\alpha$-Pinene example observes four of the
five chemical species, leaving one latent state. The NF-\(\kappa\)B benchmark uses
ten directly observed state variables and propagates uncertainty to the remaining
latent components.

\begin{table}[!t]
\caption{Summary of benchmark case studies. The table reports the number of unknown parameters
($n_\theta$), state variables ($n_x$), and state variables directly observed ($n_{\rm obs}$).}
\label{tab:benchmarks}
\centering
\begin{tabular}{lccc}
\toprule
System & $n_\theta$ & $n_x$ & $n_{\rm obs}$ \\
\midrule
Linear cascade (2-state) & 2  & 2  & 1 \\
Linear cascade (3-state) & 3  & 3  & 1 \\
Lotka--Volterra          & 4  & 2  & 1 \\
SIR                      & 2  & 3  & 1 \\
$\alpha$-Pinene          & 5  & 5  & 4 \\
NF-$\kappa$B             & 29 & 15 & 10 \\
\bottomrule
\end{tabular}
\end{table}

\subsection{Linear cascade diagnostic examples}
\label{sec:linearcascade}

Two linear cascade systems serve as analytic verification cases for the
numerical consistency of the weighted least-squares fitting, FIM covariance
estimation, and hidden-state delta-method uncertainty propagation. Because these
models admit closed-form trajectory solutions, they allow the toolbox output to
be compared against independent analytic references without the confound of
numerical ODE integration error.

\subsubsection*{Two-state cascade (LinearCascade)}

The two-state diagnostic uses the irreversible linear cascade
\begin{align}
\dot{x}_1 &= -k_1 x_1, \label{eq:lc2_x1}\\
\dot{x}_2 &= k_1 x_1 - k_2 x_2, \label{eq:lc2_x2}
\end{align}
with initial condition $x(0)=(10,0)^\top$. The synthetic benchmark uses
$(k_1,k_2)=(0.35,0.12)$. Only the downstream state $x_2$ is observed, while
$x_1$ is latent.

An independent analytic covariance reference was constructed from the
closed-form trajectory solution and compared with the FIM covariance and
hidden-state standard deviations produced by the toolbox. This checks the
Jacobian calculation, residual weighting, and delta-method propagation on a case
where the ground truth is known exactly.

\subsubsection*{Three-state cascade (LinearCascade3)}

The three-state diagnostic extends this construction to
\begin{align}
\dot{x}_1 &= -k_1 x_1, \label{eq:lc3_x1}\\
\dot{x}_2 &= k_1 x_1 - k_2 x_2, \label{eq:lc3_x2}\\
\dot{x}_3 &= k_2 x_2 - k_3 x_3, \label{eq:lc3_x3}
\end{align}
with initial condition $x(0)=(10,0,0)^\top$. The synthetic benchmark uses
$(k_1,k_2,k_3)=(0.45,0.16,0.055)$. Only the terminal state $x_3$ is observed,
while $x_1$ and $x_2$ are latent.

This gives a more demanding configuration: three unknown parameters, only the
most downstream state observed, and two latent upstream states. Observing only
the terminal species while the upstream dynamics remain hidden tests whether the
FIM and HybridCov machinery behaves sensibly when the measured signal is only
indirectly related to the hidden states. The toolbox output was compared against
a matrix-exponential reference solution.

Simulation-based calibration was also performed using a shared-fit design, so
that FIM and HybridCov received identical parameter estimates, LOO ensembles,
residual variance estimates, and sensitivity matrices in each replicate. This
ensures that FIM and HybridCov are evaluated from the same fitted branch in each
replicate, so method differences are not caused by different optimizer outcomes.
Detailed reproduction instructions are provided in Additional file 1.

\begin{figure}[H]
\centering
\includegraphics[width=0.72\linewidth]{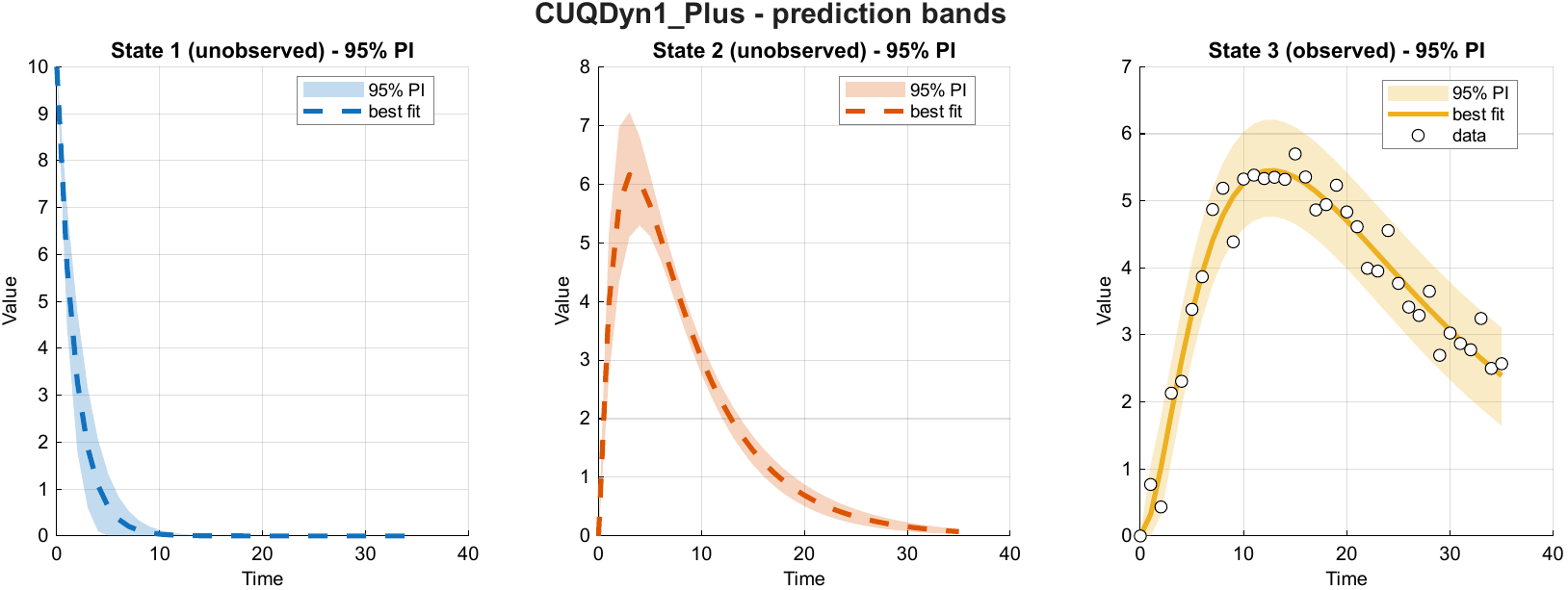}
\caption{FIM prediction bands for the three-state linear cascade (LinearCascade3) diagnostic example. The observed terminal state $x_3$ is shown with data points and conformal bands; the two hidden upstream states $x_1$ and $x_2$ are shown with delta-method bands overlaid on the analytic reference trajectory (dashed). The near-zero relative errors in Table~\ref{tab:linearref} confirm local numerical agreement with the analytic reference for this single known-truth check.}
\label{fig:lc_both}
\end{figure}

Table~\ref{tab:linearref} reports the agreement between the analytic reference and the toolbox output for both cascade configurations. All errors are at or below $10^{-6}$ in relative norm, confirming the numerical consistency of the Jacobian computation, residual weighting, and delta-method propagation on cases where the exact ground truth is known. Figure~\ref{fig:lc_both} shows the FIM prediction bands for the three-state cascade, overlaid on the analytic reference trajectories.

\begin{table}[ht]
\centering
\caption{Known-truth numerical reference checks for the linear cascade diagnostic examples. Errors are reported as the Frobenius relative error of the estimated covariance matrix, the relative error of the hidden-state standard deviation, and the maximum relative trajectory error, all computed against the analytic reference.}
\label{tab:linearref}
\begin{tabular}{lccc}
\toprule
Example & Cov.\ rel.\ Fro.\ error & Hidden std rel.\ error & Max traj.\ rel.\ error \\
\midrule
LinearCascade (2-state)  & $6.2 \times 10^{-8}$ & $1.3 \times 10^{-8}$ & $3.0 \times 10^{-8}$ \\
LinearCascade3 (3-state) & $1.4 \times 10^{-6}$ & $1.4 \times 10^{-6}$ & $8.9 \times 10^{-7}$ \\
\bottomrule
\end{tabular}
\end{table}

\begin{figure}[H]
\centering
\includegraphics[width=0.78\linewidth]{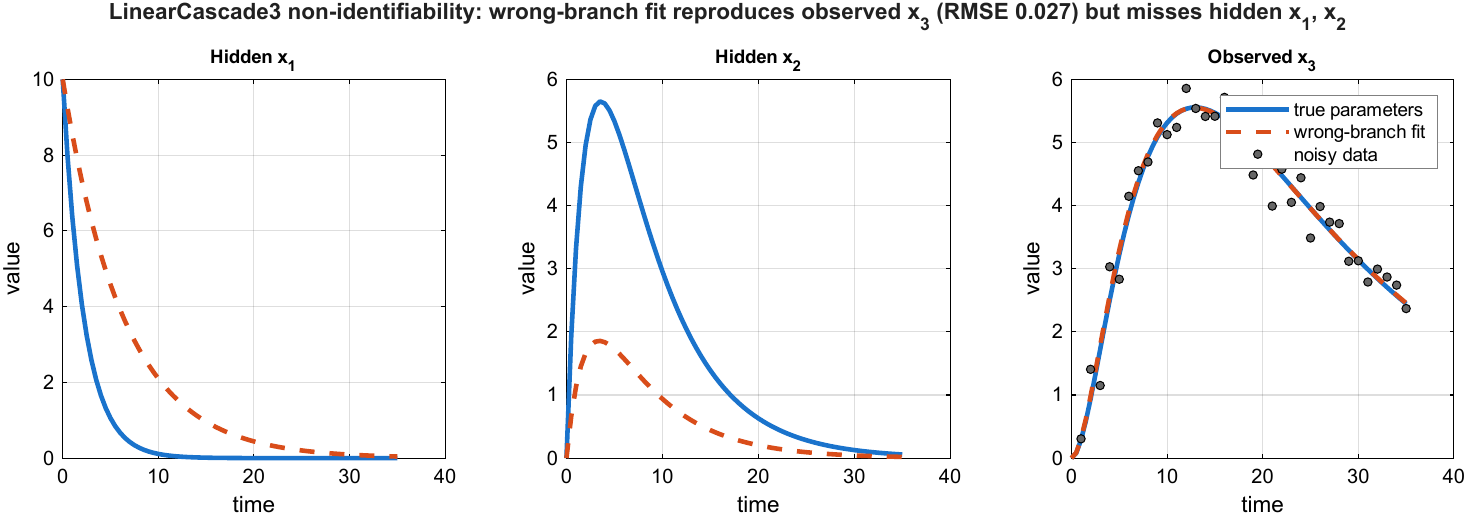}
\caption{\rev{LinearCascade3 upstream non-identifiability. A swapped-branch fit reproduces the observed terminal state $x_3$ but gives different hidden $x_1$ and $x_2$ trajectories, causing hidden-state bands to be centered on the wrong branch.}}
\label{fig:lc3_nonident}
\end{figure}

Table~\ref{tab:lc3sbc} summarizes the SBC hidden-state coverage for LinearCascade3. \rev{The analytic single-run reference confirms local numerical consistency, but the full SBC experiment exposes a separate global failure mode: because the observed terminal state depends on symmetric functions of the cascade poles, the inverse problem has a $k_1\!\leftrightarrow\!k_2$ branch symmetry. Fits on the swapped branch reproduce the observed $x_3$ trajectory nearly exactly while centering the hidden $x_1$ and $x_2$ bands on the wrong upstream trajectories (Fig.~\ref{fig:lc3_nonident}). Consequently, the mean pointwise coverage is only about 74\% at nominal 95\%, despite a well-conditioned local FIM at either branch.}

\begin{table}[ht]
\centering
\caption{SBC hidden-state coverage for LinearCascade3 (nominal 95\%, 50 replicates, shared-fit design). The full SBC run reveals global branch-switching non-identifiability that is not detected by the local FIM conditioning diagnostics.}
\label{tab:lc3sbc}
\begin{tabular}{llccc}
\toprule
Method & State & Mean ptwise cov.\ (\%) & Simult.\ cov.\ (\%) & Mean width \\
\midrule
FIM       & $x_1$ (hidden) & 73.8 & 68 & 0.670 \\
FIM       & $x_2$ (hidden) & 74.6 & 68 & 0.762 \\
HybridCov & $x_1$ (hidden) & 73.8 & 68 & 0.670 \\
HybridCov & $x_2$ (hidden) & 73.2 & 68 & 0.751 \\
\bottomrule
\end{tabular}
\end{table}

\subsection{Lotka--Volterra predator--prey system}
\label{sec:lv}

The Lotka--Volterra system \citep{wangersky1978lotka} is the main introductory benchmark and provides the most direct illustration of the partial observation setting. The two-species predator--prey model is governed by
\begin{align}
\dot{x}_1 &= x_1(\alpha - \beta x_2), \label{eq:lv1}\\
\dot{x}_2 &= -x_2(\gamma - \delta x_1), \label{eq:lv2}
\end{align}
where $x_1$ and $x_2$ represent prey and predator populations, respectively. The parameters $\alpha, \beta, \gamma, \delta$ are positive interaction constants. Synthetic datasets are generated with initial conditions $(x_1(0), x_2(0)) = (10, 5)$ and true parameters $\alpha = \gamma = 0.5$, $\beta = \delta = 0.02$.

In the partially observed setting, only the prey population $x_1$ is observed after $t_0$, while the predator $x_2$ is treated as a latent state for which prediction bands are required. This setup is representative of practical ecological and biological systems where only certain populations can be easily counted or measured.

The LV benchmark directly contrasts the three UQ workflows: conformal observed-state bands for $x_1$, and FIM delta-method, HybridCov delta-method, and bootstrap trajectory bands for $x_2$. A representative run with $B = 50$ bootstrap replicates produced the coverage and band-width results shown in Table~\ref{tab:lv_results}. All methods achieved full coverage for the corresponding known trajectories in this run. For the latent predator state $x_2$, the FIM and HybridCov delta-method bands were slightly narrower than the bootstrap trajectory band. For the observed prey state $x_1$, however, the bootstrap trajectory band was much sharper than the conformal observed-state band, because the two bands target different quantities: the conformal band includes noisy-observation prediction spread, whereas the bootstrap band shown here reflects uncertainty in the latent ODE trajectory. Thus, the width comparisons should be interpreted state by state and according to the target of each band.

\begin{table}[ht]
\centering
\caption{Representative LV benchmark results. Coverage is assessed against the known true trajectory of each state (observed prey and latent predator). Bootstrap bands are latent ODE trajectory bands, not noisy-observation prediction intervals, and should not be compared against the spread of the noisy measurements.}
\label{tab:lv_results}
\begin{tabular}{llccc}
\toprule
Method & State & Latent coverage (\%) & Mean interval width & Norm. mean width \\
\midrule
FIM          & Prey     & 100 & 14.97 & 0.194 \\
FIM          & Predator & 100 & 13.35 & 0.179 \\
HybridCov    & Prey     & 100 & 14.97 & 0.194 \\
HybridCov    & Predator & 100 & 13.40 & 0.179 \\
Bootstrap    & Prey     & 100 & 2.44  & 0.032 \\
Bootstrap    & Predator & 100 & 14.22 & 0.190 \\
\bottomrule
\end{tabular}
\end{table}

We also compared FIM and HybridCov from the same optimizer solution for this LV example. In this stricter comparison, both methods achieved similar predator coverage (approximately 93\% at nominal 95\%), confirming that for well-conditioned systems the two approaches can give comparable results. Table~\ref{tab:lvsbc} provides the detailed SBC summary.

\begin{table}[ht]
\centering
\caption{SBC predator-state coverage for Lotka--Volterra (nominal 95\%, 50 replicates, shared-fit design).}
\label{tab:lvsbc}
\begin{tabular}{lccc}
\toprule
Method & Mean ptwise cov.\ (\%) & Simult.\ cov.\ (\%) & Mean width \\
\midrule
FIM       & 93.7 & 74 & 16.12 \\
HybridCov & 93.8 & 72 & 16.58 \\
\bottomrule
\end{tabular}
\end{table}

\begin{figure}[H]
\centering
\includegraphics[width=0.88\linewidth]{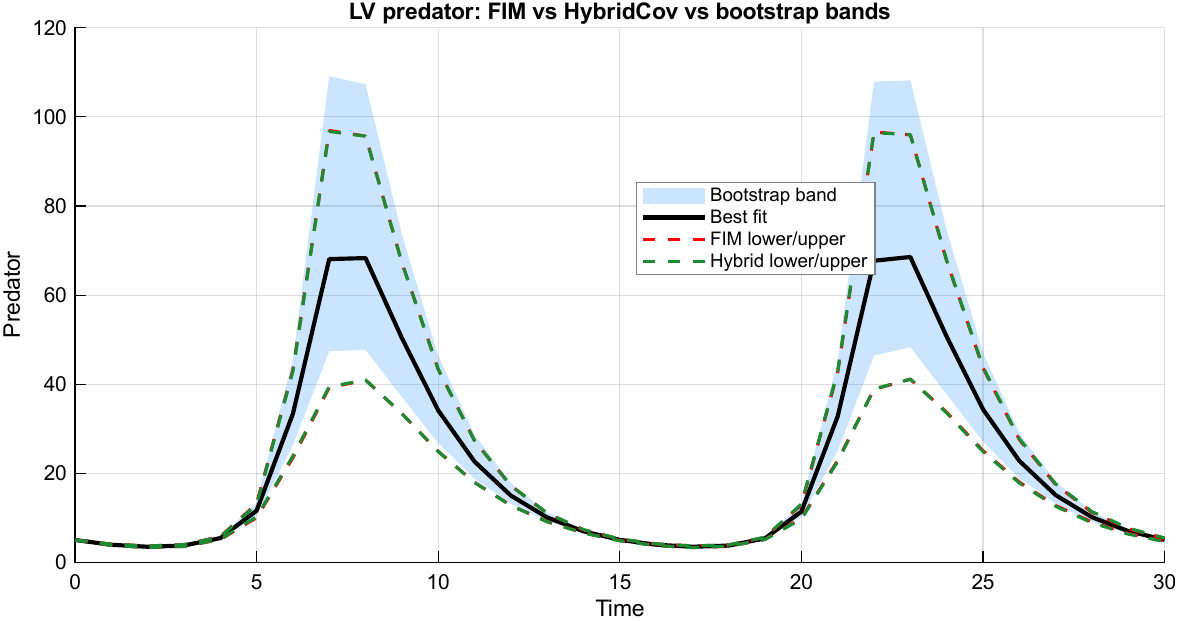}
\caption{Lotka--Volterra three-method uncertainty comparison for the latent predator ($x_2$). Shown are the parametric bootstrap band (shaded) and the FIM and HybridCov delta-method band edges (red and green dashed) around the fitted predator trajectory (solid). The FIM and HybridCov envelopes are nearly identical (mean widths 13.35 and 13.40) and overlap almost exactly, so the red FIM lines are largely hidden beneath the green HybridCov lines. }
\label{fig:lv_three}
\end{figure}

Figure~\ref{fig:lv_three} contrasts the three latent-state workflows for the predator ($x_2$): the FIM and HybridCov delta-method bands are nearly identical in width (13.35 vs 13.40) and both enclose the known true trajectory (Table~\ref{tab:lv_results}), while the parametric bootstrap band is comparable but slightly wider. The conformal band for the observed prey ($x_1$) covers the noisy measurements at the nominal level (Fig.~\ref{fig:lv_pymc}).

\subsection{SIR epidemiological model}
\label{sec:sir}

The SIR model \citep{kermack1927contribution} describes disease spread through a
population divided into susceptible ($S$), infected ($I$), and recovered ($R$)
compartments. The implementation used in this benchmark adopts the mass-action
incidence form
\begin{align}
\dot{S} &= -\beta S I, \label{eq:sir1}\\
\dot{I} &= \beta S I - \gamma I, \label{eq:sir2}\\
\dot{R} &= \gamma I, \label{eq:sir3}
\end{align}
where \(N = S + I + R\) is the fixed total population, \(\beta\) is the
mass-action transmission-rate parameter, and \(\gamma\) is the recovery rate.
For this parameterization, the initial reproductive number is
\(R_0=\beta S_0/\gamma\), or approximately \(R_0=\beta N/\gamma\) when
\(S_0 \approx N\). In the partially observed setting, only the infected
compartment \(I(t)\) is observed after \(t_0\), while \(S\) and \(R\) are latent.

This example illustrates the practical situation in epidemiology where only reported infections are available, yet predictions for susceptible and recovered compartments are important for intervention planning. The benchmark highlights parameter correlation effects ($\beta$ and $\gamma$ are strongly correlated through the basic reproduction number $R_0$) and the impact of that correlation on latent-state uncertainty propagation. The HybridCov method is expected to benefit from the empirical LOO correlation in this setting.

Table~\ref{tab:sir_results} summarizes the single-run trajectory UQ results. The conformal band for observed $I(t)$ achieves 100\% coverage with a normalized mean width of 0.10. The delta-method bands for the latent compartments $S$ and $R$ also fully cover the known true trajectories, with mean widths of 9.85 and 9.05 (FIM) and 8.64 and 9.17 (HybridCov). This is much narrower than the infected-state band on an absolute scale, as expected given that $S$ and $R$ are tightly constrained by the conservation law $S+I+R=N$ once $I(t)$ is observed. 

The SBC calibration (Table~\ref{tab:sirsbc}, nominal 90\%) reveals that empirical pointwise coverage for the latent states remains below nominal. HybridCov modestly improves mean pointwise coverage relative to FIM, from 85.0\% to 87.5\% for $S$ and from 86.6\% to 87.2\% for $R$. However, this improvement is small, and simultaneous coverage is mixed: it decreases for $S$ but increases for $R$. Thus, HybridCov does not remove the main limitation of the local Gaussian approximation for this nonlinear system; rather, it provides a limited correction to the FIM covariance structure.

Two features of the SIR system make it particularly challenging for the delta-method. First, $\beta$ and $\gamma$ are strongly correlated through the basic reproduction number $R_0 = \beta N / \gamma$: the observed infected trajectory $I(t)$ constrains $R_0$ tightly but leaves $\beta$ and $\gamma$ individually poorly determined. The FIM captures this correlation approximately, but the iso-$R_0$ manifold in parameter space is curved, so the Gaussian approximation underestimates the range of trajectories consistent with the data. Second, the latent compartments $S(t)$ and $R(t)$ respond to the \emph{long-run} integral of $I(t)$, amplifying any underestimation of parameter uncertainty into larger prediction errors than are visible in the observed-state bands alone.

This interpretation is supported by the \texttt{PyMC} ABC-SMC comparison (details below): \texttt{PyMC} gives a mean hidden-state band width of 17.6, compared with 9.45 for CUQDyn FIM and 8.90 for CUQDyn HybridCov, reflecting the broader posterior spread along the $R_0$ manifold captured by global sampling. The HybridCov strategy replaces the FIM correlation structure with an empirical LOO estimate, but since removing a single time point shifts $\beta$ and $\gamma$ only slightly, the LOO ensemble still undersamples the curved iso-$R_0$ manifold; consequently HybridCov gives only modest coverage improvement over FIM for this system.

\begin{figure}[H]
\centering
\includegraphics[width=0.72\linewidth]{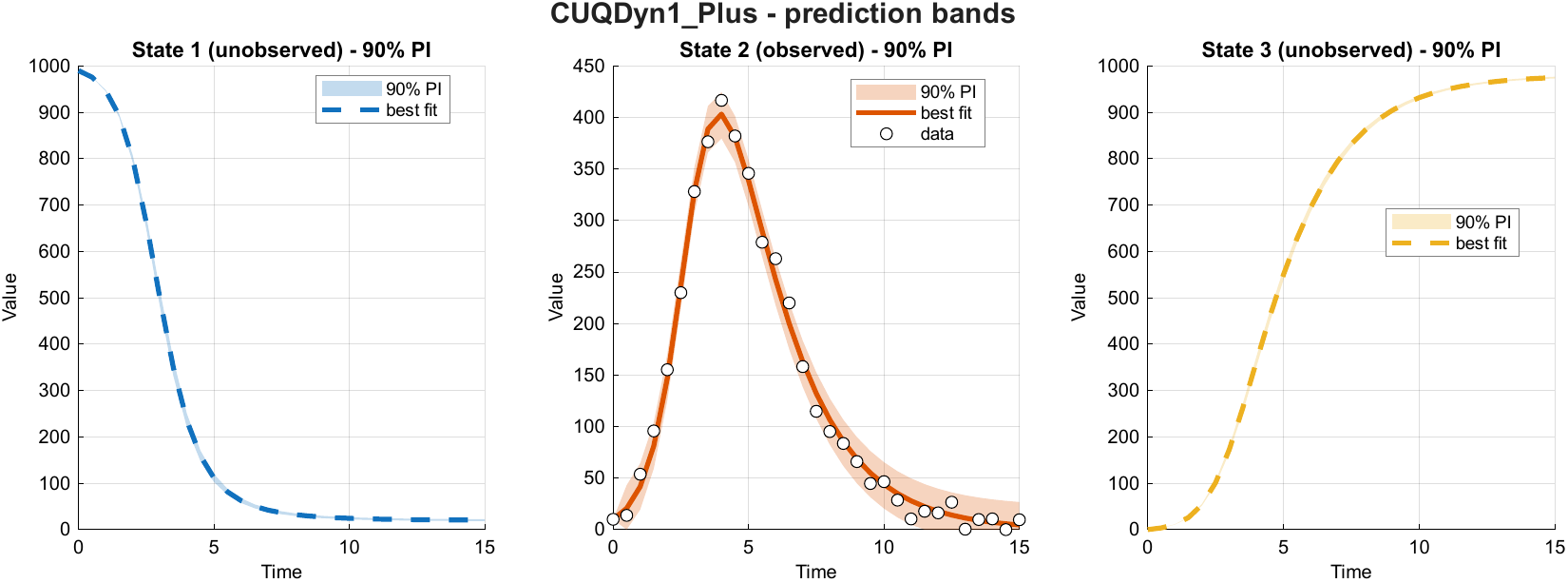}
\caption{SIR FIM prediction bands (90\% nominal). Shaded regions are prediction bands; solid lines are the fitted trajectories; dashed curves are the known true trajectories. The conformal band is applied to the observed infected compartment $I(t)$; FIM delta-method bands are applied to the latent susceptible $S(t)$ and recovered $R(t)$ compartments.}
\label{fig:sir}
\end{figure}

\begin{figure}[H]
\centering
\includegraphics[width=0.72\linewidth]{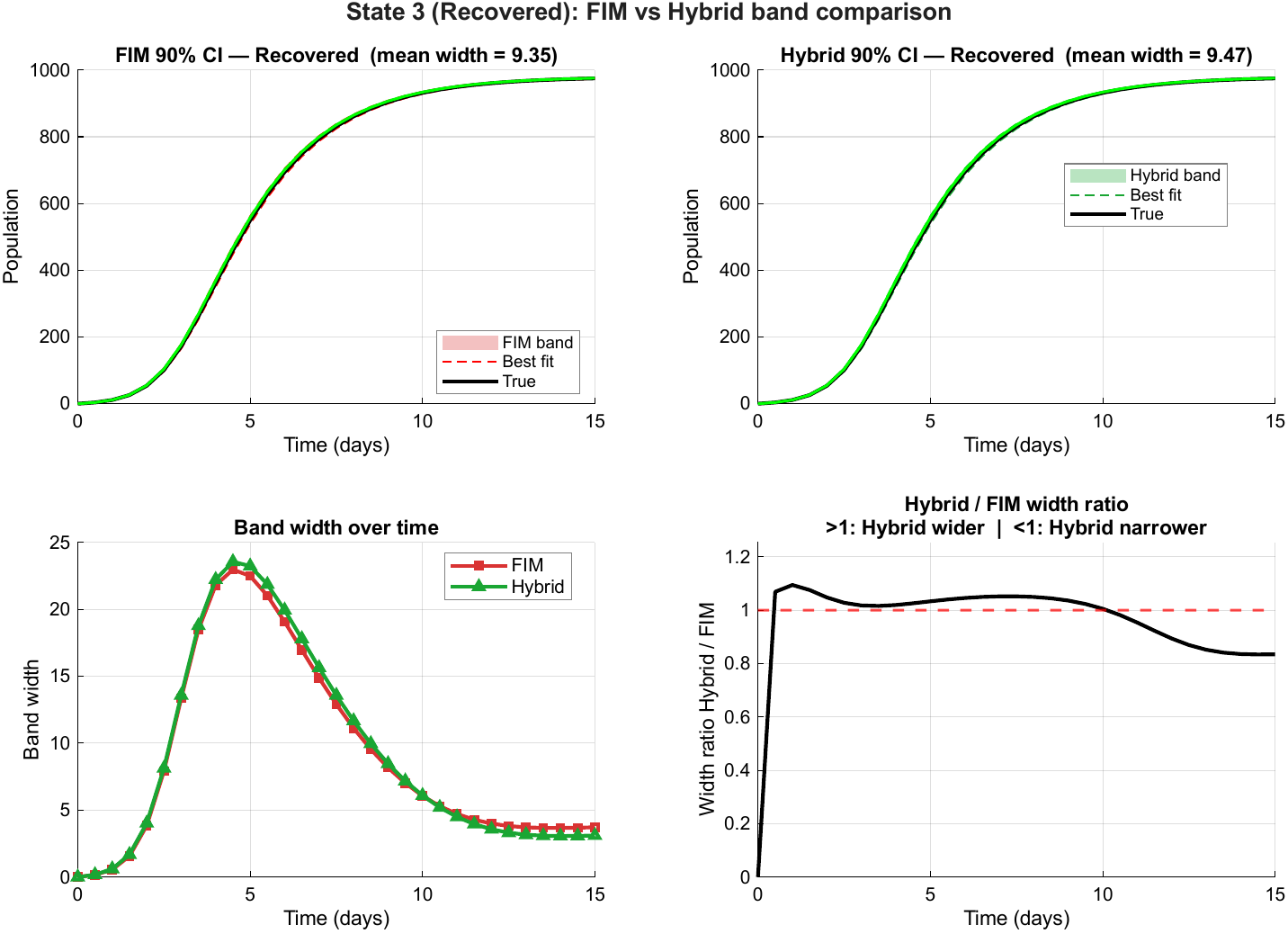}
\caption{SIR HybridCov diagnostic band comparison for the recovered state $R(t)$. The narrow delta-method band relative to the known true trajectory illustrates the under-coverage documented in Table~\ref{tab:sirsbc}, arising from underestimation of parameter uncertainty along the curved $R_0 = \beta N/\gamma$ manifold in parameter space.}
\label{fig:sir_hyb_diag}
\end{figure}

For users who require reliable latent-state coverage in SIR-type systems, the toolbox provides two practical remedies. The parametric bootstrap (Section~\ref{sec:bootstrap}) relaxes the linearity assumption by refitting the model to synthetic datasets and can capture some of the nonlinear parameter correlation, at the cost of $B$ additional model refits. Alternatively, the companion \texttt{PyMC} workflow provides a global posterior reference; where the bootstrap and \texttt{PyMC} bands agree and are substantially wider than the delta-method bands, this is strong evidence that delta-method bands should be used with caution for those latent states. Figure~\ref{fig:sir} shows the FIM prediction bands; Figure~\ref{fig:sir_hyb_diag} shows the HybridCov diagnostic comparison for the recovered state.

\begin{table}[ht]
\centering
\caption{SIR single-run trajectory UQ summary (90\% nominal level). $I$ is the conformal observed-state band; $S$ and $R$ are delta-method bands for latent states. Coverage is assessed against the known true trajectory. Normalized widths use the conserved total population $N \approx 418$ as the reference scale, consistent across all three compartments.}
\label{tab:sir_results}
\begin{tabular}{llccc}
\toprule
Method & State & Coverage (\%) & Mean width & Norm.\ width \\
\midrule
FIM       & $S$ (latent)   & 100 &  9.85 & 0.024 \\
FIM       & $I$ (observed) & 100 & 43.48 & 0.104 \\
FIM       & $R$ (latent)   & 100 &  9.05 & 0.022 \\
HybridCov & $S$ (latent)   & 100 &  8.64 & 0.021 \\
HybridCov & $I$ (observed) & 100 & 43.48 & 0.104 \\
HybridCov & $R$ (latent)   & 100 &  9.17 & 0.022 \\
\bottomrule
\end{tabular}
\end{table}

\begin{table}[ht]
\centering
\caption{SBC latent-state coverage for SIR (nominal 90\%, 50 replicates). Both FIM and HybridCov fall short of the nominal level, suggesting limitations of the local Gaussian approximation for this nonlinear system.}
\label{tab:sirsbc}
\begin{tabular}{llccc}
\toprule
Method & State & Mean ptwise cov.\ (\%) & Simult.\ cov.\ (\%) & Mean width \\
\midrule
FIM       & $S$ (Susceptible) & 85.0 & 80 & 10.098 \\
FIM       & $R$ (Recovered)   & 86.6 & 72 & 9.287 \\
HybridCov & $S$ (Susceptible) & 87.5 & 78 & 9.991 \\
HybridCov & $R$ (Recovered)   & 87.2 & 78 & 9.252 \\
\bottomrule
\end{tabular}
\end{table}

\subsection{$\alpha$-Pinene isomerization}
\label{sec:alpinene}

The $\alpha$-pinene isomerization model \citep{box1973problems} describes the thermal rearrangement of $\alpha$-pinene into several isomeric products through a network of five coupled reactions:
\begin{align}
\dot{x}_1 &= -(p_1 + p_2) x_1, \label{eq:ap1}\\
\dot{x}_2 &= p_1 x_1, \label{eq:ap2}\\
\dot{x}_3 &= p_2 x_1 - (p_3 + p_4) x_3 + p_5 x_5, \label{eq:ap3}\\
\dot{x}_4 &= p_3 x_3, \label{eq:ap4}\\
\dot{x}_5 &= p_4 x_3 - p_5 x_5. \label{eq:ap5}
\end{align}

\rev{The example observes only $x_1$--$x_4$ after the initial time and treats $x_5$ as hidden. Following the original $\alpha$-pinene data of \citet{box1973problems}, the four measured isomer time series are labelled $y_1$--$y_4$ (the notation also used in Additional file~1); under the identity observation map used here, $y_k$ is the noisy measurement of state $x_k$ for $k=1,\ldots,4$ (Eq.~\ref{eq:noise}), so the observed-state set is $\mathcal{O}=\{1,2,3,4\}$. Thus $x_1$--$x_4$ receive observed-state jackknife$+$-style empirical bands, while $x_5$ receives FIM or HybridCov delta-method latent-state bands.}

\begin{figure}[H]
\centering
\includegraphics[width=0.72\linewidth]{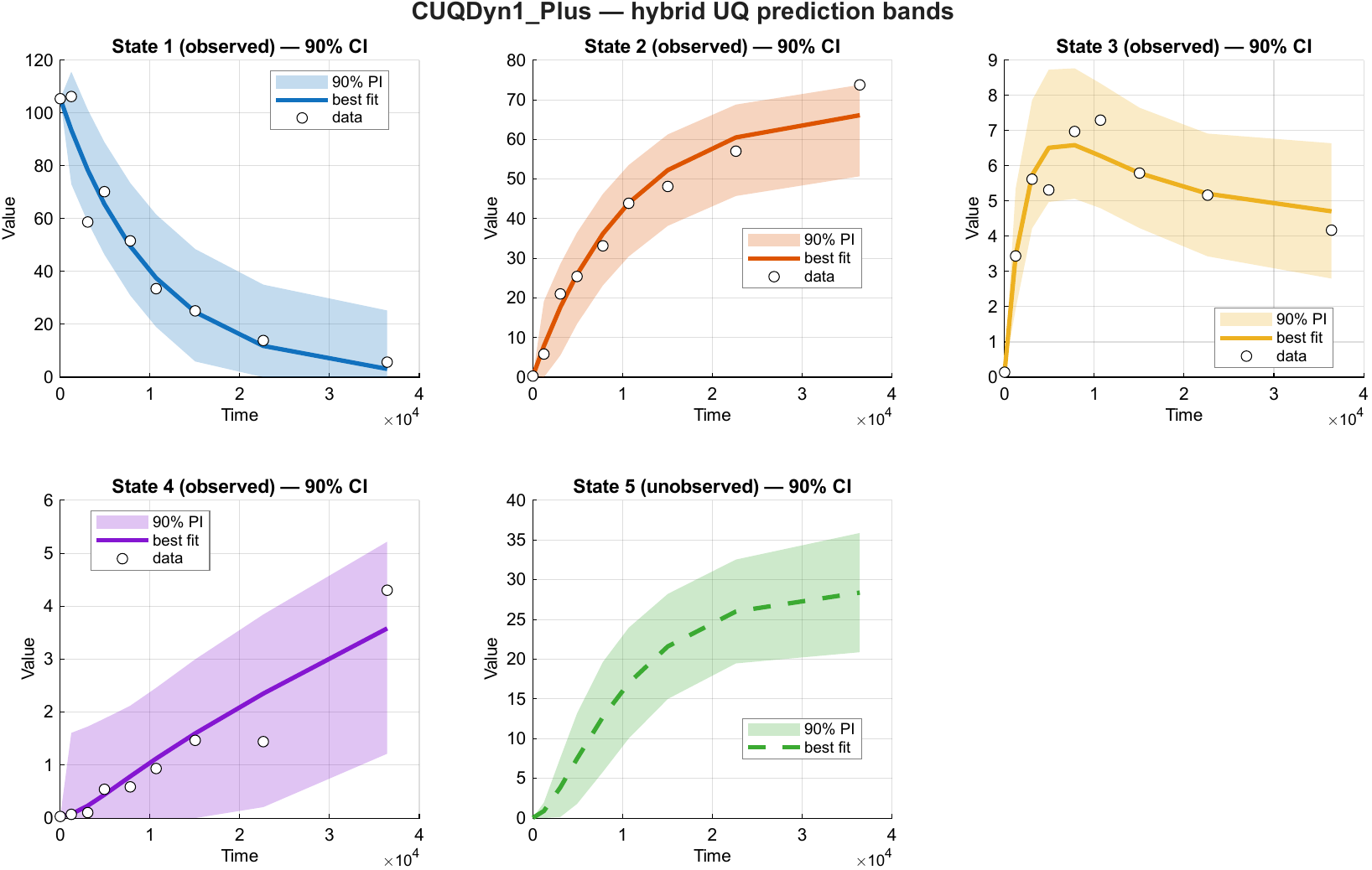}
\caption{$\alpha$-Pinene isomerization HybridCov prediction bands (90\% nominal) for the maintained partially observed configuration. States $x_1$--$x_4$ use observed-state empirical conformal bands; hidden state $x_5$ uses a HybridCov delta-method latent-state band. Open circles are observations; dashed curves show the reference trajectories where available.}
\label{fig:ap}
\end{figure}

\rev{Table~\ref{tab:ap_results} summarizes the AP trajectory comparison for the \texttt{CUQDyn1\_Plus} workflows. The observed states achieve 100\% empirical coverage in this run, and the hidden $x_5$ mean band width is approximately 9.93 under FIM and 10.0 under HybridCov. SBC calibration (Table~\ref{tab:apsbc}) at nominal 90\% gives near-nominal but slightly under-covering hidden-state bands: 88.5\% for FIM and 87.2\% for HybridCov. Figure~\ref{fig:ap} shows the HybridCov UQ bands, with conformal bands on $x_1$--$x_4$ and a latent delta-method band on $x_5$.}
\begin{table}[ht]
\centering
\caption{$\alpha$-Pinene single-run trajectory UQ summary for the maintained partially observed configuration (90\% nominal level). States $x_1$--$x_4$ are observed and use empirical conformal bands; $x_5$ is hidden and uses delta-method latent-state bands. Coverage and RMSE are reported only for observed states.}
\label{tab:ap_results}
\begin{tabular}{llrrrr}
\toprule
State & Type & FIM mean width & HybridCov mean width & Obs. cov. (\%) & Obs. RMSE \\
\midrule
$x_1$ & observed & 37.8 & 37.8 & 100 & 8.17 \\
$x_2$ & observed & 20.5 & 20.5 & 100 & 3.54 \\
$x_3$ & observed & 3.20 & 3.20 & 100 & 0.569 \\
$x_4$ & observed & 2.95 & 2.95 & 100 & 0.403 \\
$x_5$ & hidden   & 9.93 & 10.0 & --- & --- \\
\bottomrule
\end{tabular}
\end{table}

\begin{table}[ht]
\centering
\caption{SBC coverage for $\alpha$-pinene state $x_5$ (nominal 90\%, 50 replicates).}
\label{tab:apsbc}
\begin{tabular}{lccc}
\toprule
Method & Mean ptwise cov.\ (\%) & Simult.\ cov.\ (\%) & Mean width \\
\midrule
FIM       & 88.5 & 78 & 4.015 \\
HybridCov & 87.2 & 80 & 3.958 \\
\bottomrule
\end{tabular}
\end{table}

\subsection{NF-$\kappa$B signaling pathway}
\label{sec:nfkb}

The NF-$\kappa$B signaling pathway \citep{lipniacki2004mathematical} is the most computationally challenging benchmark, with 15 state variables and 29 unknown parameters (ODEs are not given here for the sake of brevity). The system describes the regulatory dynamics of the Nuclear Factor Kappa B transcription factor in response to inflammatory stimuli. Using the conventional NF-\(\kappa\)B state notation \(y_i\), ten selected
state variables are directly observed and the remaining five state variables
are treated as latent. This model-specific notation follows the original
NF-\(\kappa\)B formulation; elsewhere in the article, state variables are
denoted generically by \(x_i\).

This example represents a typical large-scale systems biology problem where high dimensionality, partial observability, and potential parameter non-identifiability make Bayesian MCMC approaches impractical. In our original study \citep{portela2025conformal}, Stan failed to converge to adequate posterior predictive regions for the NF-$\kappa$B system even after many hours of computation. The CUQDyn1 and CUQDyn2 conformal algorithms computed prediction regions for all ten observables within minutes.

In the \texttt{CUQDyn1\_Plus} setting, the NF-$\kappa$B example demonstrates the scalability of the LOO conformal workflow to high-dimensional nonlinear systems and the utility of FIM and HybridCov delta-method propagation for the five unobserved latent states. The toolbox diagnostic functions allow inspection of covariance conditioning and parameter identifiability.

Table~\ref{tab:nfkb_results} summarizes the aggregate \texttt{CUQDyn1\_Plus} trajectory UQ results. Conformal bands for the ten observed states achieve mean coverage of 96.2\% with FIM and 96.8\% with HybridCov, at a nominal 90\% level. The mean observed-state RMSE is 2.15 (FIM) and 2.16 (HybridCov), dominated by $y_{11}$ whose large absolute scale inflates the mean. For the five hidden states, the mean delta-method band width is very large, especially under HybridCov, reflecting ill-conditioned sensitivity--covariance products in this high-dimensional system. These entries should be interpreted together with the FIM weak-direction reliability diagnostics rather than as precise quantitative uncertainty intervals; per-state details and the \texttt{PyMC} comparison are provided in Additional file 1 and Section~\ref{sec:pymc_comparison}.

\begin{table}[ht]
\centering
\caption{NF-$\kappa$B aggregate \texttt{CUQDyn1\_Plus} trajectory UQ summary (90\% nominal for observed-state bands). Hidden-state widths are local delta-method quantities and should be interpreted with the FIM reliability diagnostics.}
\label{tab:nfkb_results}
\small
\begin{tabular}{lrrrr}
\toprule
Method & Obs.\ coverage (\%) & Obs.\ RMSE & Obs.\ width & Hidden width \\
\midrule
CUQDyn FIM       & 96.2 &  2.15 &  8.68 & $4.993\times10^{4}$ \\
CUQDyn HybridCov & 96.8 &  2.16 & 10.1  & $7.151\times10^{4}$ \\
\bottomrule
\end{tabular}

\end{table}
The toolbox completes the full LOO ensemble and UQ computation within minutes on a standard workstation, in contrast to the convergence failures encountered by Stan on this system in our earlier study \citep{portela2025conformal}. Supplementary CUQDyn1\_Plus--PyMC comparison results and implementation details are provided in Additional file 1. The two workflows produce visually similar bands for the well-identified observables; slight differences appear for the more poorly constrained \rev{outputs}, consistent with the expectation that the HybridCov empirical correlation provides a modest correction over the pure FIM approximation in weakly identifiable directions.

\section{Comparison with Bayesian Inference}
\label{sec:pymc_comparison}

We compare \texttt{CUQDyn1\_Plus} with a Bayesian reference implemented via approximate Bayesian computation with sequential Monte Carlo (ABC-SMC) in \texttt{PyMC} \cite{pymc2023}. The comparison uses the same benchmark datasets for both approaches in the benchmark systems. The \texttt{PyMC} analyses set explicit random seeds for reproducibility. For LV, the comparison run uses 4000 ABC-SMC draws per chain across 4 chains; SIR, $\alpha$-pinene, and the tabulated NF-$\kappa$B comparison use 1000 draws per chain across 4 chains. All \texttt{PyMC} trajectory summaries are computed from 500 posterior trajectory draws. For observed states, \texttt{PyMC} bands include an additive observation-noise term estimated from the posterior-mean RMSE; hidden-state \texttt{PyMC} bands remain latent parameter-uncertainty bands.\texttt{CUQDyn1\_Plus}hidden-state bands are local FIM or HybridCov delta-method bands.

Sampler convergence is assessed with the rank-normalized potential scale reduction factor $\hat{R}$ \citep{gelman1992inference,vehtari2021rank} and the bulk effective sample size (ESS) \citep{vehtari2021rank}; we regard a run as converged when $\hat{R}\le 1.01$ for all parameters and the bulk-ESS is at least $100$ per chain, and flag it otherwise. The NF-$\kappa$B \texttt{PyMC} workflow uses a broad uniform prior over the same parameter support used by CUQDyn, $[0.1\,\theta_{\mathrm{nom}},4\,\theta_{\mathrm{nom}}]$, rather than a log-normal prior centred on nominal values. This change removes the previous concern that the Bayesian prior encoded the answer, but it also exposes the weak identifiability of the 29-parameter NF-$\kappa$B problem: the posterior remains close to the broad-prior mean for many parameters. 

The tabulated NF-$\kappa$B comparison uses 1000 draws per chain and does not meet the convergence criterion ($\hat{R}\le 1.01$ and effective sample size $\ge 100$ per chain); an independent 4000-draw sensitivity run improved the diagnostics to a marginal pass ($\hat{R}\approx 1.01$, minimum bulk ESS $\approx 503$) but did not remove the prior-dominated parameter-recovery pattern. The NF-$\kappa$B \texttt{PyMC} entries are therefore reported as a not-fully-converged, prior-dominated reference rather than a trusted posterior, and convergence and identifiability must be interpreted separately.


Table~\ref{tab:param_comparison} summarizes the latest parameter-recovery comparison. LV and SIR are well-identified synthetic examples: \texttt{CUQDyn1\_Plus} and \texttt{PyMC} both recover all parameters to within 20\% relative error and agree closely in point estimates. In AP, which is based on experimental data rather than synthetic truth, \texttt{CUQDyn1\_Plus} remains closer to the nominal reference values, while \texttt{PyMC} drifts more strongly for $p_4$ and $p_5$ under the broader prior support. The trajectory-level comparison is nevertheless much closer than the raw parameter errors suggest, consistent with parameter compensation under partial observation. NF-$\kappa$B is the stress test: \texttt{CUQDyn1\_Plus} recovers a subset of parameters accurately but reports very large covariance scales in weak directions, whereas \texttt{PyMC} remains prior-dominated. NF-$\kappa$B rate constants should therefore not be presented as reliably recovered by either method.

\begin{table}[ht]
\centering
\caption{Parameter recovery summary for the CUQDyn1\_Plus--PyMC comparison. Relative errors are computed against synthetic truth for LV, SIR, and NF-$\kappa$B, and against nominal/reference values for $\alpha$-pinene.}
\label{tab:param_comparison}
\begin{tabular}{llccc}
\toprule
Model & Method & $n_\theta$ & Mean relative error (\%) & Parameters $\leq 20\%$ \\
\midrule
\multirow{3}{*}{LV}
  & CUQDyn FIM       & 4  &  4.8 & 4/4 \\
  & CUQDyn HybridCov & 4  &  4.8 & 4/4 \\
  & PyMC ABC-SMC     & 4  &  6.2 & 4/4 \\
\midrule
\multirow{3}{*}{SIR}
  & CUQDyn FIM       & 2  &  0.7 & 2/2 \\
  & CUQDyn HybridCov & 2  &  0.7 & 2/2 \\
  & PyMC ABC-SMC     & 2  &  0.6 & 2/2 \\
\midrule
\multirow{3}{*}{$\alpha$-Pinene}
  & CUQDyn FIM       & 5  & 23.0 & 3/5 \\
  & CUQDyn HybridCov & 5  & 23.0 & 3/5 \\
  & PyMC ABC-SMC     & 5  & 97.0 & 2/5 \\
\midrule
\multirow{3}{*}{NF-$\kappa$B}
  & CUQDyn FIM       & 29 & 45.2 & 14/29 \\
  & CUQDyn HybridCov & 29 & 54.7 & 10/29 \\
  & PyMC ABC-SMC     & 29 & 107.4 & 0/29 \\
\bottomrule
\end{tabular}
\end{table}

For LV,\texttt{CUQDyn1\_Plus}and \texttt{PyMC} recover $\alpha$, $\beta$, $\delta$, and $\gamma$ with nearly identical estimates. For example, CUQDyn FIM gives $\alpha=0.507\pm0.023$ and \texttt{PyMC} gives $0.511\pm0.028$; for $\gamma$,\texttt{CUQDyn1\_Plus}gives $0.483\pm0.035$ and \texttt{PyMC} gives $0.482\pm0.042$. SIR is even more tightly aligned:\texttt{CUQDyn1\_Plus}and \texttt{PyMC} both estimate $\beta\approx0.001981$ and $\gamma\approx0.499$.

For AP, the strongest discrepancies are $p_4$ and $p_5$. \texttt{CUQDyn1\_Plus}  estimates $p_4=3.67\times10^{-4}\pm2.94\times10^{-4}$ and $p_5=5.93\times10^{-5}\pm9.01\times10^{-5}$, whereas \texttt{PyMC} gives $p_4=7.80\times10^{-4}\pm2.98\times10^{-4}$ and $p_5=1.33\times10^{-4}\pm4.00\times10^{-5}$. The large\texttt{CUQDyn1\_Plus}standard deviations and \texttt{PyMC} posterior drift both point to weak practical identifiability rather than a simple numerical disagreement.


Table~\ref{tab:band_comparison} compares trajectory-band behavior. LV and SIR again show close agreement in observed-state coverage and RMSE. For AP, \texttt{PyMC} and \texttt{CUQDyn1\_Plus} give similar observed-state RMSE and high coverage, despite larger parameter disagreement. NF-$\kappa$B differs qualitatively: \texttt{CUQDyn1\_Plus} gives much lower observed-state RMSE in the tabulated run, while \texttt{PyMC} obtains high coverage through much wider observed-state bands after noise augmentation. Hidden-state NF-$\kappa$B bands must be interpreted together with FIM conditioning and weak-direction diagnostics.

\begin{table}[ht]
\centering
\caption{Trajectory-band comparison for the CUQDyn1\_Plus--PyMC comparison. Observed-state \texttt{PyMC} bands include estimated observation noise; hidden-state \texttt{PyMC} bands are posterior trajectory quantiles. The NF-kB \texttt{PyMC} row derives from a 1000-draw run that does not meet the $\hat{R}$/ESS convergence criterion and should be read as a prior-dominated reference.}
\label{tab:band_comparison}
\scriptsize
\setlength{\tabcolsep}{2pt}
\begin{tabular}{llrrrr}
\toprule
Model & Method & Obs.\ cov.\ (\%) & Obs.\ RMSE & Obs.\ width & Hidden width \\
\midrule
\multirow{3}{*}{LV}
  & CUQDyn FIM       & 100.0 & 2.33 & 15.0 & 13.3 \\
  & CUQDyn HybCov    & 100.0 & 2.33 & 15.0 & 13.4 \\
  & PyMC ABC-SMC     &  96.8 & 2.58 & 10.9 & 15.7 \\
\midrule
\multirow{3}{*}{SIR}
  & CUQDyn FIM       & 100.0 & 10.0 & 43.5 & 9.45 \\
  & CUQDyn HybCov    & 100.0 & 10.0 & 43.5 & 8.90 \\
  & PyMC ABC-SMC     & 100.0 & 10.0 & 43.6 & 17.6 \\
\midrule
\multirow{3}{*}{$\alpha$-Pinene}
  & CUQDyn FIM       & 100.0 & 3.17 & 16.1 & 9.93 \\
  & CUQDyn HybCov    & 100.0 & 3.17 & 16.1 & 10.0 \\
  & PyMC ABC-SMC     &  97.2 & 3.62 & 18.7 & 20.4 \\
\midrule
\multirow{3}{*}{NF-$\kappa$B}
  & CUQDyn FIM       & 96.2 &  2.15 &  8.68 & $4.993\times10^{4}$ \\
  & CUQDyn HybCov    & 96.8 &  2.16 & 10.1 & $7.151\times10^{4}$ \\
  & PyMC ABC-SMC     & 98.1 & 19.0  & 88.4  & 442 \\
\bottomrule
\end{tabular}
\end{table}

\begin{figure}[H]
\centering
\includegraphics[width=\linewidth]{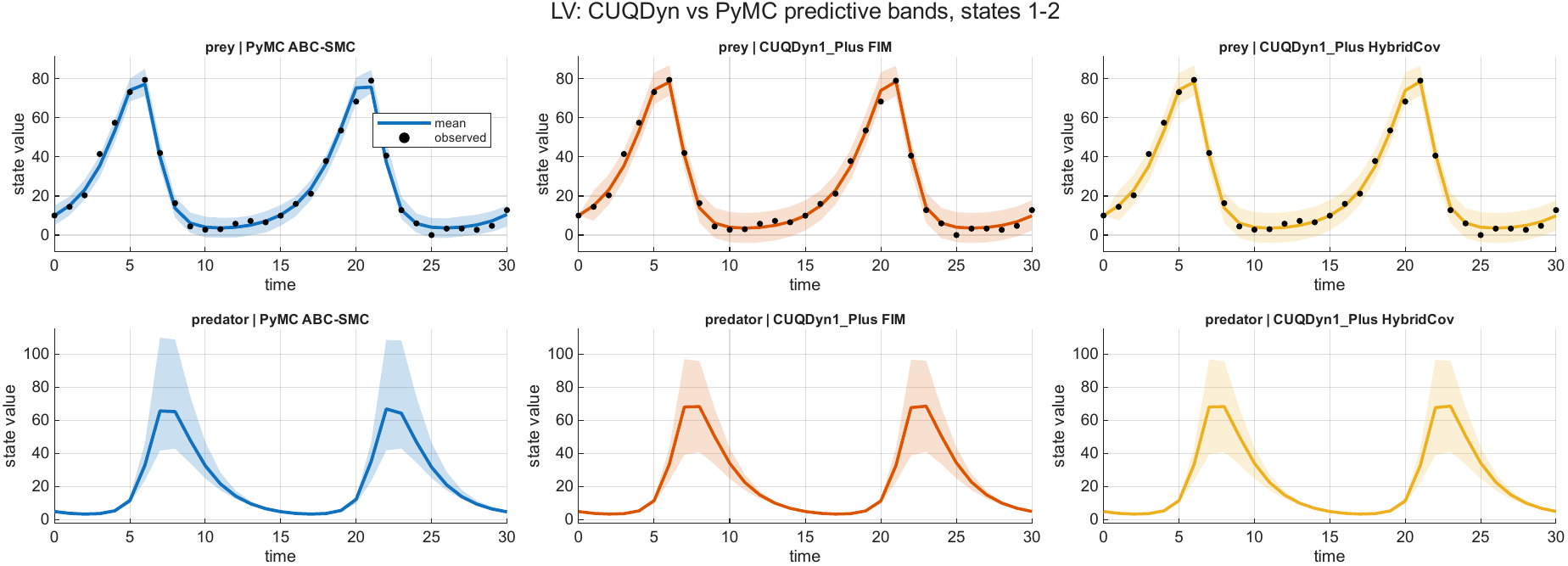}
\caption{Lotka--Volterra side-by-side comparison: \texttt{PyMC} ABC-SMC (left), CUQDyn FIM (centre), and CUQDyn HybridCov (right). Top row: observed prey ($x_1$); bottom row: latent predator ($x_2$).}
\label{fig:lv_pymc}
\end{figure}
\begin{figure}[H]
\centering
\includegraphics[width=\linewidth]{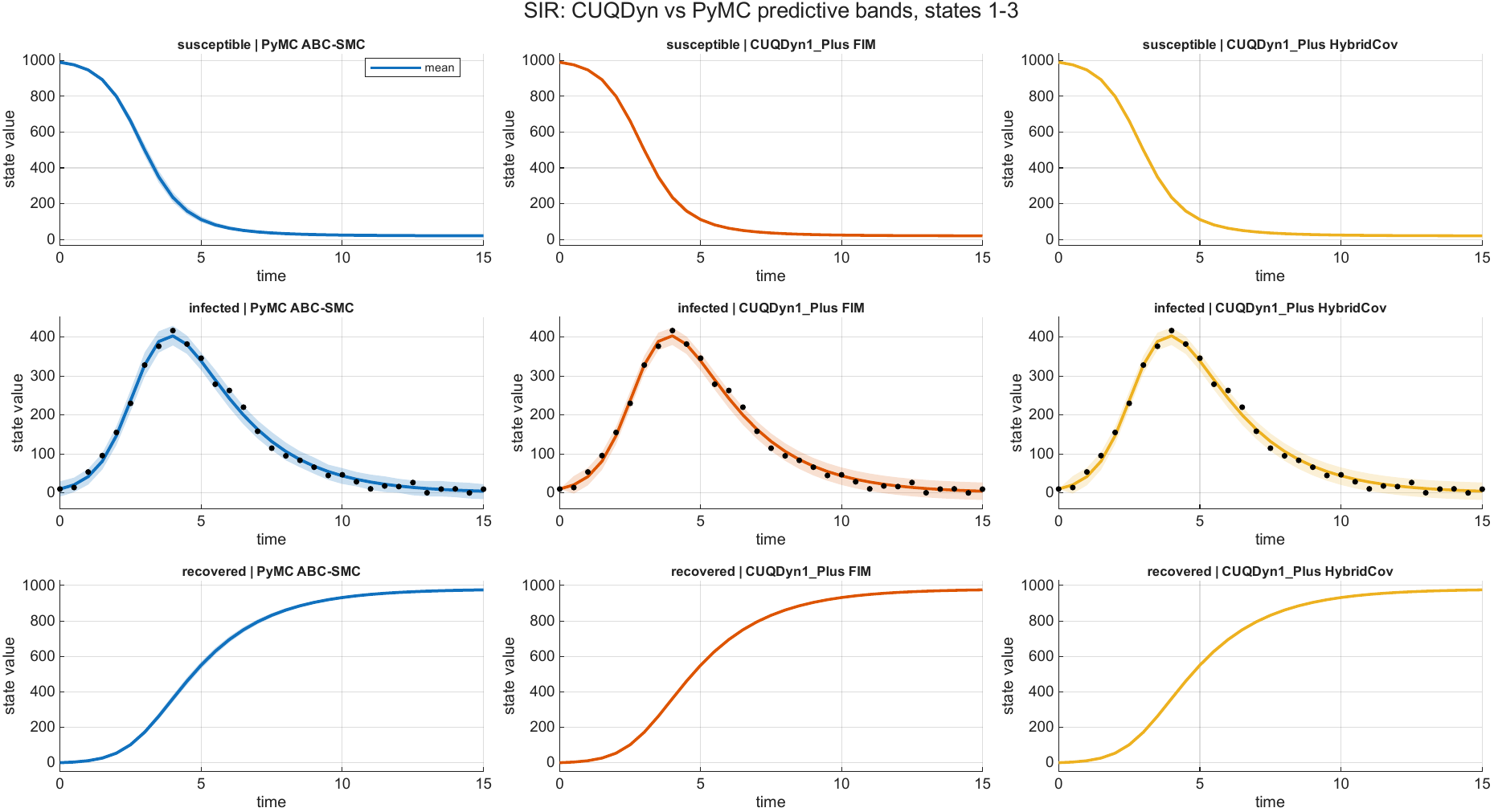}
\caption{SIR side-by-side comparison: \texttt{PyMC} ABC-SMC (left), CUQDyn FIM (centre), and CUQDyn HybridCov (right). Rows correspond to susceptible, infected, and recovered states.}
\label{fig:sir_pymc}
\end{figure}
\begin{figure}[H]
\centering
\includegraphics[width=\linewidth]{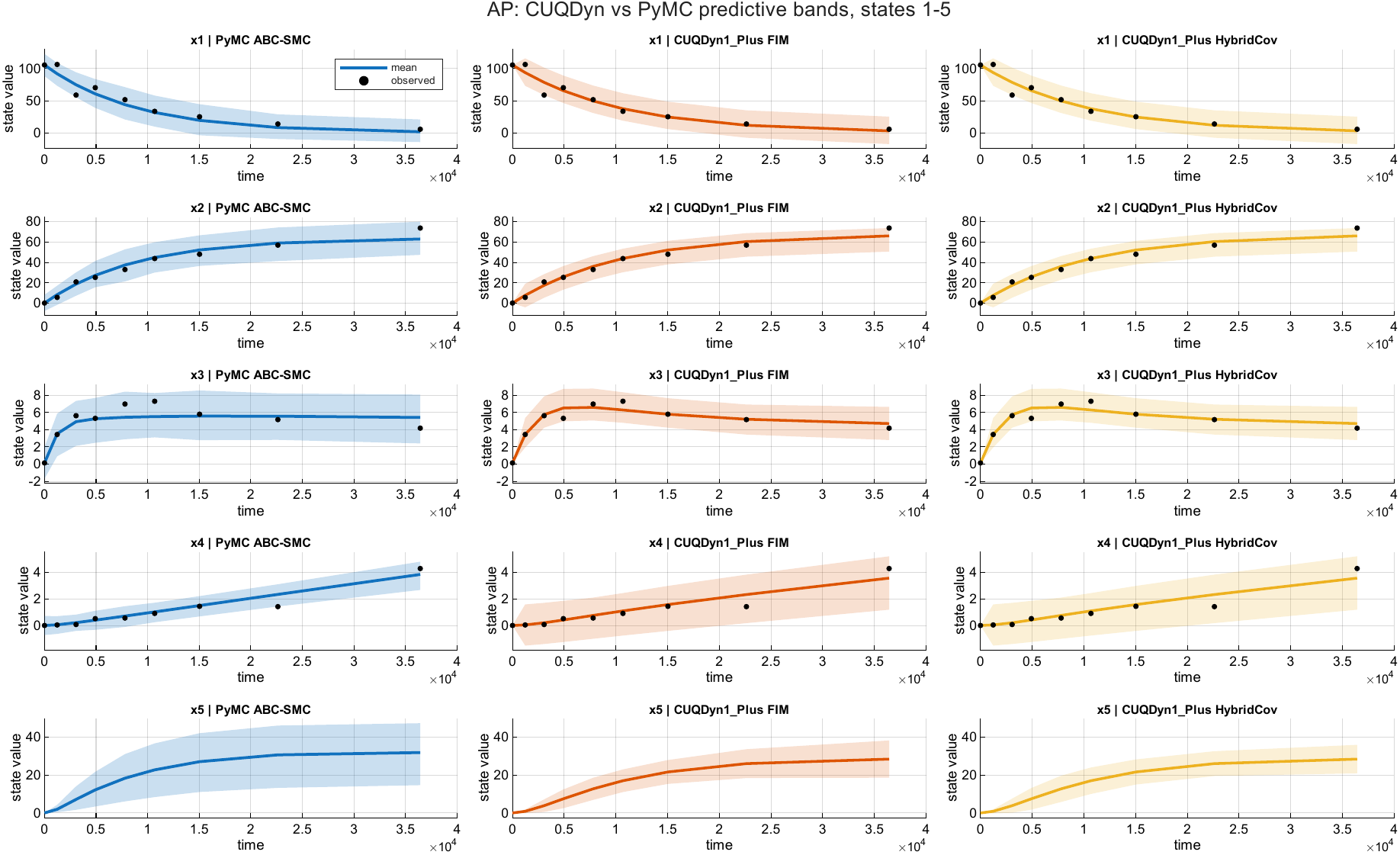}
\caption{$\alpha$-Pinene side-by-side comparison: \texttt{PyMC} ABC-SMC (left), CUQDyn FIM (centre), and CUQDyn HybridCov (right). State $x_5$ is hidden after the initial condition.}
\label{fig:ap_pymc}
\end{figure}

Figures~\ref{fig:lv_pymc}--\ref{fig:ap_pymc} show side-by-side panels for \texttt{PyMC} ABC-SMC, CUQDyn FIM, and CUQDyn HybridCov. Within each state row, the y-axis limits are identical across methods. For the NF-\(\kappa\)B case study, the full state-by-state \texttt{CUQDyn1Plus-PyMC} predictive comparison is provided in Additional file 1. We omit these panels from the main text because the 15-state comparison is visually too dense, and because the main conclusion is better conveyed by the aggregate trajectory and parameter summaries in Tables~\ref{tab:param_comparison} and~\ref{tab:band_comparison}: NF-\(\kappa\)B remains weakly identifiable for both workflows, with \texttt{PyMC} posterior predictive bands reflecting broad uncertainty and \texttt{CUQDyn1\_Plus}  FIM-based bands requiring reliability-diagnostic interpretation for weakly informed latent directions.

For LV and SIR, the visual agreement is close for both observed and hidden states. For AP, the observed states are tracked comparably by all methods, while the hidden $y_5$ band is wider under PyMC, consistent with posterior spread in weakly constrained parameters. For NF-$\kappa$B, the broad-prior \texttt{PyMC} posterior remains prior-dominated and gives very wide observed bands for some states, particularly around $y_{11}$ dynamics. \texttt{CUQDyn1\_Plus} identifies a good data-fitting region and produces tighter observed-state bands, but the extremely large hidden-state delta-method bands for $y_8$ and $y_{10}$ demonstrate why FIM reliability diagnostics are required before interpreting latent-state uncertainty in this high-dimensional, partially observed model.

The comparison supports a nuanced conclusion. \texttt{CUQDyn1\_Plus} and \texttt{PyMC} agree well on LV and SIR, where the inverse problems are well posed. AP shows that predictive agreement can be stronger than parameter agreement when partially observed dynamics allow compensation among weakly identified parameters. NF-$\kappa$B shows that neither broad-prior ABC-SMC nor local Gaussian covariance propagation should be interpreted as full rate-constant recovery in a high-dimensional, weakly identifiable model. In this regime, the most reliable comparison is based on observed trajectory fit, predictive coverage, convergence diagnostics, FIM conditioning, and per-state latent-band reliability diagnostics rather than on raw parameter intervals alone.

\color{black}
\section{Discussion}
\label{sec:discussion}

We have presented an uncertainty quantification framework, and its software implementation (\texttt{CUQDyn1\_Plus}), which can handle partially observed dynamic (ODE-based) models of biological systems. This framework extends the conformal prediction algorithms of \citet{portela2025conformal} to the practically important setting where only a subset of model states are experimentally observable. This method combines leave-one-out jackknife$+$-style empirical calibration for observed states with local Gaussian uncertainty propagation for latent states.

The key design principle is the asymmetric treatment of observed and unobserved states, which reflects the fundamentally different statistical information available for each type of variable. For observed states, the leave-one-out conformal construction calibrates prediction bands directly from held-out residuals, \rev{providing conformal-style empirical calibration without imposing Gaussian residuals.} For unobserved states, no calibration data exist, and uncertainty must be propagated analytically through the ODE sensitivity structure. \rev{The FIM delta-method approach now uses a shared log-space, rank-aware covariance helper and stores reliability diagnostics for weak directions;} the HybridCov strategy modifies this baseline by incorporating empirical parameter correlation information from the LOO ensemble.

The simulation-based calibration experiments reported here are an important validation step that goes beyond point coverage checks on single datasets. By repeatedly generating synthetic data, fitting, and evaluating coverage, SBC provides direct evidence about whether the nominal coverage level is achieved in the model and data regime of interest. \rev{The results should be described as model-dependent rather than as a uniform HybridCov win: naive LOO-only covariance is under-dispersed, while FIM-only and HybridCov performance depends on conditioning, nonlinearity, and the informativeness of the LOO correlation estimate.}

The proposed framework provides several practical advantages relative to existing approaches. Its cost scales as $\mathcal{O}(m^2 n_\theta k)$ for the LOO ensemble (Section~\ref{sec:complexity}), avoiding the $10^4$--$10^5$ posterior samples per chain typical of MCMC. In our earlier study this amounted to up to two orders of magnitude less computation than Stan for equivalent problems \citep{portela2025conformal}, and in the present work the full estimation-and-UQ workflow returns usable prediction bands in minutes on every benchmark, including the NF-$\kappa$B model on which the \texttt{PyMC} samplers did not converge. We deliberately do not report a controlled wall-clock comparison here: \texttt{CUQDyn1\_Plus} is a MATLAB research prototype, whereas the PyMC/ABC-SMC backend compiles its most expensive operations, so raw run times would reflect implementation maturity and language as much as algorithmic cost. The efficiency argument therefore rests on algorithmic scaling and on the ability to obtain a usable result without certifying sampler convergence, rather than on a like-for-like timing benchmark. An optimized implementation and a matched runtime study are natural future work.

Another advantage of our approach is that its observed-state bands provide jackknife$+$-style calibration that is less dependent on a correctly specified parametric noise model, a meaningful robustness property when the exact measurement error distribution is unknown. Its compatibility with arbitrary nonlinear ODE models through user-defined dynamics and cost functions enables straightforward extension to new biological systems. Because it relies on optimization and analytic propagation rather than posterior sampling, our workflow also avoids the convergence and mixing failures that MCMC and SMC samplers commonly encounter on high-dimensional, weakly identifiable models, illustrated here by the non-converged ABC-SMC run on NF-$\kappa$B and, previously, by Stan on the same system \citep{portela2025conformal}. Our approach delivers reproducible results in minutes, while its diagnostics make the remaining identifiability limits explicit rather than hiding them in an unconverged posterior.

Several limitations of the current framework should be acknowledged. Most fundamentally, uncertainty quantification for hidden states still relies on local Gaussian approximations. The delta method assumes that ODE trajectories vary approximately linearly with parameters near $\hat{\theta}$ and that parameter uncertainty is approximately Gaussian. These assumptions are likely to fail for strongly nonlinear systems, multimodal likelihood surfaces, weakly identifiable parameter combinations, and models with chaotic or discontinuous dynamics. The bootstrap workflow provides a partial remedy by relaxing the linearity assumption, but it is more computationally expensive and introduces its own assumptions about the bootstrap noise model. The SIR benchmark illustrates this mode in the present study: empirical coverage of latent states is below nominal 90\%, and the delta-method bands are approximately half the width of the \texttt{PyMC} posterior predictive bands, pointing to underestimation of uncertainty along the curved $R_0 = \beta N/\gamma$ manifold in parameter space. \rev{The LinearCascade3 SBC experiment reveals a complementary failure mode: a global $k_1\!\leftrightarrow\!k_2$ branch symmetry can produce well-conditioned local FIM covariances around the wrong parameter branch, so local diagnostics alone are insufficient to guarantee hidden-state coverage. A related and practically important failure mode is weak identifiability in high-dimensional systems such as NF-$\kappa$B. The current log-space FIM implementation improves conditioning and reports SVD rank diagnostics, but it cannot remove structural non-identifiability. Hidden states whose sensitivities project onto weak FIM directions are therefore flagged as unreliable, and their bands should be interpreted as regularization-dependent local approximations rather than definitive uncertainty statements.}

Second, the leave-one-out conformal calibration can become computationally expensive for datasets with many time points or expensive ODE models. The quadratic scaling in the number of time points ($\mathcal{O}(m^2)$ in the worst case) limits the approach for dense time series, though parallelization and the local-after-global LOO refit strategy substantially mitigate this in practice.
Third, conformal coverage guarantees apply directly only to observed states at observed time points. Extending distribution-free guarantees to hidden-state trajectories or to future unobserved time points remains an important open research problem. Approaches based on conformal risk control \citep{angelopoulos2023conformal} or trajectory-level conformal sets may offer pathways toward distribution-free guarantees for latent states, though their application to ODE systems with partial observation would require substantial methodological development.

Looking ahead, several extensions would broaden the applicability of the framework. Scalable approximate LOO methods based on influence functions or approximate leave-one-out cross-validation \citep{luo2023iterative} could reduce the $\mathcal{O}(m^2)$ computational bottleneck. Integration with optimal experimental design frameworks could use the sensitivity and covariance information computed by the toolbox to guide measurement selection for reducing latent-state uncertainty. Hybrid Bayesian--conformal frameworks that use the conformal calibration as a posterior predictive correction could combine the strengths of both approaches. 

The \texttt{CUQDyn1\_Plus} codebase, examples, and SBC workflows are publicly available and are designed to be extended by users developing new ODE models. The high-level problem definition interface, automated code generation, and standardized output reporting are intended to make the toolbox accessible to systems biologists who are not specialists in conformal prediction or optimization, while the modular source organization supports methodological extensions by advanced users.

\backmatter

\bmhead{Acknowledgements}
The authors thank the developers and maintainers of the external scientific software used in the benchmark workflows.

\section*{Declarations}

\bmhead{Funding}
JRB acknowledges support from grant PID2023-146275NB-C22 (DYNAMO-bio) funded by MICIU/AEI/10.13039/501100011033 and ERDF/EU, and from grant CSIC PIE 202470E108 (LARGO).

\bmhead{Competing interests}
The authors declare that they have no competing interests.

\bmhead{Ethics approval and consent to participate}
Not applicable.


\bmhead{Availability of data and materials}
The datasets used in the benchmark examples are included with the toolbox repository or generated by the scripts described in the manuscript.

\bmhead{Code availability}
\texttt{CUQDyn1\_Plus} is implemented in MATLAB and released under the \rev{GPLv3} license. The software, including instructions for running all experiments, is available at Zenodo: \href{https://doi.org/10.5281/zenodo.21470609}{https://doi.org/10.5281/zenodo.21470609}.

\bmhead{Additional files}
\textbf{Additional file 1.} File format: PDF. Title: Supplementary information for \texttt{CUQDyn1\_Plus}. Description: Detailed computational evaluation, including problem-definition and tutorial material, empirical benchmark summaries, simulation-based calibration results, \texttt{CUQDyn1\_Plus-PyMC} ABC-SMC comparison tables and figures, identifiability and HybridCov diagnostics, Bayesian convergence diagnostics, per-run settings, method notes, and scope limitations.

\bmhead{Authors' contributions}
AP and JRB contributed to conceptualization, methodology, software, validation, formal analysis, and writing--review and editing. JRB wrote the original draft, supervised the work, and acquired funding. Both authors read and approved the final manuscript.

\bibliography{references}

\end{document}